\documentclass[12pt]{article}
\usepackage{amsmath,epsfig,subfigure,amsthm,amsfonts}

\usepackage[sort]{natbib}

\makeatletter \@addtoreset{equation}{section}

\newcommand{\beq}[1]{\begin{equation} \label{#1}}
\newcommand{\eeq}{\end{equation}}
\newcommand{\bed}{\begin{displaymath}}
\newcommand{\eed}{\end{displaymath}}
\newcommand{\ben}{\begin{eqnarray*}}
\newcommand{\een}{\end{eqnarray*}}

\def\qed{\hfill{$\Box$}}

\def\cd{(\cdot)}
\def\M{{\cal M}}
\def\rr{{\mathbb R}}

\newcommand{\var}{\text{Var}}

\newcommand{\dl}{\delta}
\newcommand{\e}{\varepsilon}

\newcommand{\al}{\alpha}

\newcommand{\sg}{\sigma}
\newcommand{\pr}{{P}}
\newcommand{\ex}{E}
\newcommand{\wdt}{\widetilde}
\newcommand{\wdh}{\widehat}

\def\para#1{\vskip 0.4\baselineskip\noindent{\bf #1}}

\newtheorem{thm}{Theorem}[section]
\newtheorem{prop}[thm]{Proposition}
\newtheorem{lem}[thm]{Lemma}

\theoremstyle{definition}
\newtheorem{rem}[thm]{Remark}
\newtheorem{defn}[thm]{Definition}
\newtheorem{exm}[thm]{Example}

\newcommand{\thmref}[1]{Theorem~{\rm \ref{#1}}}
\newcommand{\lemref}[1]{Lemma~{\rm \ref{#1}}}

\newcommand{\propref}[1]{Proposition~{\rm \ref{#1}}}

\newcommand{\exmref}[1]{Example~{\rm \ref{#1}}}
\newcommand{\figref}[1]{Figure~{\rm \ref{#1}}}

\newcommand{\set}[1]{\left\{#1\right\}}

\newcommand{\disp}{\displaystyle}

\newcommand{\bedd}{\bed\begin{array}{l}}
\newcommand{\eedd}{\end{array}\eed}

\def\({\left(}
\def\){\right)}
\def\l{\left|}
\def\r{\right|}
\newcommand{\nd}{\noindent}

\def\one{{\hbox{1{\kern -0.35em}1}}}

\newcommand{\bea}{\bed\begin{array}{rl}}
\newcommand{\eea}{\end{array}\eed}
\newcommand{\ad}{&\!\!\!\disp}
\newcommand{\aad}{&\disp}
\newcommand{\barray}{\begin{array}{ll}}
\newcommand{\earray}{\end{array}}
\newcommand{\lbar}{\overline}

\begin{document}

\title{Numerical Solutions of Optimal Risk
Control and Dividend Optimization
Policies under A Generalized Singular Control Formulation}

\author{Zhuo Jin,\thanks{Centre for Actuarial Studies, Department
of Economics, The University of Melbourne, VIC 3010, Australia,
zjin@unimelb.edu.au.} \and G.
Yin,\thanks{Department of Mathematics, Wayne State University,
Detroit, Michigan 48202, gyin@math.wayne.edu. 
}
 \and Chao
Zhu\thanks{Department of Mathematical Sciences, University of
Wisconsin-Milwaukee, Milwaukee, WI 53201, zhu@uwm.edu. 
}}

\maketitle

\begin{abstract}

This paper develops numerical methods for finding optimal dividend
pay-out and reinsurance policies. A generalized singular control
formulation of surplus and discounted payoff function are
introduced, where
the surplus is modeled by a regime-switching process subject to
both regular and singular controls. To approximate the value
function and optimal controls,
 Markov chain approximation
techniques are used to construct a discrete-time controlled Markov
chain with two components. The proofs of the convergence of the
approximation sequence to the surplus process and the value
function are given. Examples of proportional and excess-of-loss
reinsurance are presented to illustrate the applicability of the
numerical methods.


\vskip 0.25 in \nd{\bf Key Words.}  Singular control, dividend
policy, Markov chain approximation, numerical method, reinsurance,
regime switching.

\end{abstract}

\newpage

\section{Introduction}\label{sec:intro}

To design optimal risk controls and and dividend payout strategies
for a financial corporation has drawn increasing attention since
the introduction of the classical collective risk model in
\cite{Lundberg}, where the probability of ruin was considered as a
measure of risk. Realizing that the surplus reaching arbitrarily
high and exceeding any finite level are not realistic in practice,
\cite{Finetti} proposed an dividend optimization problem. Instead
of considering the safety aspect (ruin probability), aiming at
maximizing the expected discounted total dividends until lifetime
ruin by assuming the surplus process follows a simple random walk,
he showed that the optimal dividend strategy is a barrier
strategy. Since then, many researchers have
 analyzed this problem
under more realistic assumptions and extended its range of applications.
Some recent work can be found in
\cite{AsmussenT,ChoulliTZ,GerberS04} and references therein. To
protect insurance companies against the impact of claim
volatilities, reinsurance is a standard tool with the goal of
reducing and eliminating risk. The primary insurance carrier pays
the reinsurance company a certain part of the premiums. In return,
the reinsurance company is obliged to share the risk of large
claims. Proportional reinsurance is one type of reinsurance
policy. Within this scheme, the reinsurance company covers a fixed
percentage of losses. The other type of reinsurance policy is
nonproportional reinsurance. The most common nonproportional
reinsurance policy is the so-called excess-of-loss reinsurance,
within which the cedent (primary insurance carrier) will pay all
of the claims up to a pre-given level of amount (termed retention
level). The comparison of these two types of reinsurance can be
found in \cite{AsmussenHT}. In this paper, we  consider both of
these reinsurance policies and provide the numerical solutions of
the corresponding Markovian regime-switching models.

Let $u(t)$ be an exogenous retention level, which is a control
chosen by the insurance company representing the reinsurance
policy. In a Crem\'er-Lundberg model, claims
arrive according a Poisson process with 
 rate $\beta$. Let
$Y_i$ be the size of the $i$th claim. The $Y_i$'s are independent
and identically distributed (i.i.d.) random variables. Let
 $Y_i^u$ be the the fraction of the claims hold by the
cedent. The insurer selects the time and the amount of dividends
to be paid out to the policyholders. Let $X(t) $ denote the
controlled surplus of an insurance company at time $t\ge 0$.
Throughout this paper, we only consider cheap
reinsurance,
 where the safety loading
for the reinsurer is the same as that for the cedent. The numerical scheme and
the convergence proofs are also applicable to more general
reinsurance problems. By using the
techniques of diffusion approximation applied to the
Crem\'er-Lundberg model, the surplus process satisfies
\beq{approxLundberg}
\begin{cases}
dX(t)=\beta E[Y_i^{u}]dt+\sqrt{\beta E[(Y_i^{u})^2])} dw(t), \\
X(0^-)=x,
\end{cases}
\eeq where $w(t)$ is a standard Brownian motion. In the case of
proportional reinsurance, $Y_i^{u}=u Y_i$. Thus, following
\eqref{approxLundberg}, the surplus is given by \beq{proportion}
\begin{cases}
dX(t)=\beta u(t)E[Y_i]dt+u(t)\sqrt{\beta E[Y_i^2]} dw(t), \\
X(0^-)=x,
\end{cases}
\eeq In the case of excess-of-loss reinsurance, $Y_i^{u}=Y_i
\wedge u$ with the retention level $u$. We have \beq{moment}
\barray \ad E[Y^{u}]=\int_0^{u}\bar F(x)dx, \
E[(Y^{u})^2]=\int_0^{u} 2x \bar F(x)dx \earray \eeq where $\bar
F(x)=P(Y_i > x)$. The stochastic differential equation of the
surplus process follows \beq{nonproportion} \left \{ \barray \ad
dX(t)=\beta \int_0^{u(t)}\bar F(x)dx dt+ \Big[\beta
\int_0^{u(t)}\bar F(x)dx \Big]^{\frac{1}{2}} dw(t), \\ \ad
X(0^-)=x. \earray \right. \eeq

A common choice of the payoff is to maximize the total expected
discounted value of all dividends until lifetime ruin; see
\cite{GerberS06} and \cite{JinYY11}. Let
\beq{tau}\tau:=\inf\set{t>0: X(t)\notin G} \eeq be the ruin time,
where $G=(0,\infty)$ is the domain of the surplus. Denote by $r>0$
the discounting factor, and by $Z(t)$ the total dividends paid out
up to time $t$. Our goal is to maximize \beq{costfun1}
E\int_0^\tau e^{-rt} dZ(t). \eeq Some ``bequest" functions and
more complicated utility functions are added to the payoff
functions in the work \cite{Browne95,Browne97}. In this paper, we
treat payoff functions that are more general and complex than
those given in \eqref{costfun1} or \cite{Browne95,Browne97}; our
proposed numerical methods are easily implementable.

A dividend strategy $Z\cd$ is an ${\cal F}_t$-adapted process
$\{Z(t): t\geq0\}$ corresponding to the accumulated amount of
dividends paid up to time $t$ such that $Z(t)$ is a nonnegative
and nondecreasing stochastic process that is right continuous with
left limits. Throughout the paper, we use the convention that
$Z(0^-)=0$.
 In general, a dividend process
is not necessarily   absolutely continuous. In fact, dividends
are not usually paid out continuously in practice. For instance,
insurance companies may distribute dividends on discrete time
intervals resulting in unbounded payment rate. In such a scenario,
 the surplus level changes drastically on a dividend payday.
Thus abrupt or discontinuous changes occur due to
  ``singular'' dividend distribution policy.
Together with proportional or excess-of-loss reinsurance policy,
this gives rise to a mixed regular-singular stochastic control
problem.

Empirical studies indicate in particular that traditional surplus
models fail  to capture more extreme price movements. To better
reflect reality, much effort has been devoted to producing better
models. One of the recent trends is to use regime-switching
models. \cite{Hamilton} introduced a regime-switching time series
model. Recent work on risk models and related issues can be found
in \cite{Asmussen-89,Yang-Yin04}. In \cite{WYW10}, the optimal
dividend and proportional reinsurance strategy under utility
criteria were studied for the regime-switching compound Poisson
model by using the methods of the classical and impulse control
theory.  \cite{SotomayorC} obtained optimal dividend strategies
under a regime-switching diffusion model. A comprehensive study of
switching diffusions with ``state-dependent'' switching is in
\cite{YinZ10}.

In this work, we model the surplus process by a regime-switching
diffusion;  reinsurance and dividend payment policies are
 introduced as regular and singular stochastic controls. The goal is
 to maximize the  expected total discounted payoff until ruin;
 see \eqref{payoff-numerical} for details.
The model we consider appears to be more versatile and realistic
than the classical compound Poisson or diffusion models. To find
the optimal reinsurance and dividend pay-out strategies,
one usually  solves  a  so-called  Hamilton-Jacobi-Bellman (HJB) equation.
However, in our work, thanks to regime switching and
the mixed regular and singular control formulation, the HJB equation
is in fact a coupled system of nonlinear
 quasi-variational inequalities (QVIs).
 A closed-form solution is virtually impossible to obtain. A
viable alternative is to employ numerical approximations. In this
work, we  adapt the Markov chain approximation methodology
developed by \cite{Kushner-D}.
 To the best of our knowledge,  numerical methods for
singular controls of regime-switching diffusions have not been
studied in the literature to date. Even for singular controlled
diffusions without regime switching, the related results are
relatively scarce; \cite{Budhiraja-Ross} and \cite{Kushner-M-1991}
are the only papers  that carry out a convergence analysis  using
weak convergence and relaxed control formulation of numerical
schemes for  singular control problems in the setting of It\^o
diffusions. We focus on developing numerical methods that are
applicable to mixed regular and singular controls for
regime-switching models. Although the primary motivation stems
from insurance risk controls, the techniques and the algorithms
suggested appear to be applicable to other singular control
problems. It is also worth mentioning that the Markov chain
approximation method requires little regularity of the value
function and/or
 analytic properties of the associated systems of
HJB equations and/or QVIs. The numerical implementation can be
done using either value iterations or policy iterations.

The rest of the paper is organized as follows. A generalized
formulation of optimal risk control and dividend policies and
assumptions are presented in Section \ref{sec:form}. The two most
common types of reinsurance strategy (proportional reinsurance and
excess-of-loss reinsurance) are covered in our study. Section
\ref{sec:app-mc} deals with the numerical algorithm of Markov
chain approximation method. The regular control
 and the singular control are well approximated by the approximating Markov
chain and the dynamic programming equation are presented. Section
\ref{sec:conv} deals with the convergence of the approximation
scheme. The technique of ``rescaling time" is introduced and the
convergence theorems are proved. Two classes of numerical examples
are provided in Section \ref{sec:exm} to illustrate the
performance of the approximation method. Finally, some additional
remarks are provided in Section \ref{sec:rmk}.

\section{Formulation}\label{sec:form}
In this section, we introduce a dynamic system to describe the
surplus processes with reinsurance and dividend payout strategies
with Markov regime switching. Let $X(t) $ denote the controlled
surplus of an insurance company at time $t\ge 0$. Denote by $u(t)$
and   $Z(t)$ the dynamic reinsurance policy at time $t$ and the
total dividend paid out  up to time $t$, respectively. Assume the
evolution of $X(t)$, subject to reinsurance and dividend payments,
follows a one-dimensional temporal homogeneous controlled
regime-switching
diffusion on an   
 unbounded domain $G=(0,\infty)$:
 \beq{1d-state}
  \begin{cases}
dX(t)= b(X(t),\al(t),u(t))dt +
\sigma(X(t),\al(t),u(t))dW(t) - dZ(t),
 \\ X(0^-)=x \in G, \ \ \al(0^-)=\ell\in \M,
 \end{cases}
\eeq
where
$u$ is the regular control
 and $Z$ is the singular control.
Throughout the paper we use the convention  that $Z(0^-)=0$. The jump size of
$Z$ at time $t\ge 0$ is denoted by $\Delta Z(t):= Z(t)-Z(t^-) $, and $Z^c(t) :=
Z(t)-\sum_{0\le s \le t} \Delta Z(s)$ denotes the continuous part of $Z$.
Also note that $\Delta X(t) : =X(t)-X(t^-)= - \Delta Z(t)$ for any $t\ge 0$.

Denote by $r>0$ the discounting factor.
For  suitable functions $f$ and $c$  
and an arbitrary admissible pair $\pi=(u,Z)$,
the expected discounted payoff is
 \beq{payoff-numerical} J(x,\ell,\pi)=
 \ex_{x,\ell} \left[\int_{0}^{\tau} e^{-
rt}\left[f(X(t),\al(t),u(t))dt  +
c(X(t^-),\al(t^-))dZ(t)\right]
 \right]. \eeq
The pair $\pi=(u,Z)$ is said to be {\em admissible} if $u$ and $Z$ satisfy
 \begin{itemize}
 \item[(i)] $u(t)$ and $Z (t) $ are nonnegative for
 any $t\ge 0$,
 \item[(ii)]   $Z $ is c\`adl\`ag and nondecreasing,
 \item[(iii)]  $     X (t )\ge 0$, for any $t\le \tau$, where $\tau$ is the ruin
time defined in \eqref{tau},
 \item[(iv)] both $u$ and $Z$ are adapted to ${\cal F}_t:=\sigma\set{W(s),\al(s),
0\le s \le t}$ augmented by the $\pr$-null sets,  and
 \item[(iv)] $J(x,\ell,\pi) < \infty$ for any $(x, \ell)\in G\times \M $ and admissible pair $\pi=(u,Z)$, where $J$
is the functional defined in \eqref{payoff-numerical}.
 \end{itemize}

 Suppose that $\mathcal A $ is the collection of all admissible
 pairs, and
 $U$ is the collection of possible retention levels $u(t)$. Throughout the paper,
 we
 assume that
 $U$ is a given compact set, and that for each $\ell \in  \M$, $c(x,\ell) \ge c(y,\ell)$
 for all $0\le x\le y$. That is, the utility function for the dividend is
 non-decreasing, see examples in \cite{GerberS05} and \cite{Alvarez}.
In addition, $c(X(t),\ell)=f(X(t),\ell,u)=0$ when $t > \tau$.
 Define the value
function as
  \beq{value-numerical-purpose}
 V(x,\ell) := \sup_{\pi \in \cal A} J(x,\ell, \pi).
 \eeq

If the value function $V$ defined in
\eqref{value-numerical-purpose} is sufficiently smooth,
by applying the dynamic programming principle (\cite{FlemingS}), we conclude formerly
 that $V$ satisfies the following coupled system of quasi variational inequalities (QVIs):
 \beq{qvi}
\max \left\{
H(x,\ell,  V'(x,\ell),  V''(x,\ell)) + Q(x) V(x,\cdot) (\ell)-
rV(x,\ell), c (x,\ell)-  V'(x,\ell)  \right\} =0,
\eeq for all $ (t,x,\ell)\in [0,\tau)\times G\times \M,$ with
boundary condition \beq{boundary-data} V(0,\ell)=0, \ \ \forall
\ell\in \M,\eeq where  for any $( x,\ell,p) \in  \times \rr^n
\times \M \times \rr^n $,  \bea \ad H(x,\ell,p):= \sup_{u\in U} \big\{f( x,\ell,u)
 + p\cdot b( x,\ell,u) + \frac{1}{2} \sigma^2( x,\ell,u)\big\},
  \\
  \ad Q(x)V( x,\cdot)(\ell):= \sum_{\iota\in \M} q_{\ell \iota}(x)
   [V( x,\iota)-V( x,\ell)],
 \eea
and $V'$ and $V''$ denote the first and the second partial
derivatives of $V$ with respect to $x$.
Note that coupling in \eqref{qvi} is due to the term
 $Q(x)V(t,x,\cdot)(\ell)$, which is not contained in the  usual QVI
 (as in \cite{FlemingS,Ma-Yong,Pha09}).

Nevertheless, the value function $V$ is not necessarily smooth. In
fact, there are examples (\cite{Bayraktar}) where the value
function is not even continuous. In our work, both the ruin time
$\tau$ and the reinsurance $u$ and the dividend $Z$ policies  may
depend on the initial surplus level, which leads to nonsmooth vale
function.
   Moreover, \eqref{qvi} is a coupled system of nonlinear
   differential equations. A closed-form solution to \eqref{qvi} is
   by and large impossible. Therefore in this work, we propose
   a numerical scheme to approximate the value
   function as well as optimal reinsurance and dividend payment policies.

\section{Numerical Algorithm}\label{sec:app-mc}

Our goal is to design a numerical scheme to approximate value
function $V$ in \eqref{value-numerical-purpose}. As a standing
assumption, we assume $V\cd$ is continuous with respect to $x$.
In this section we construct a locally consistent
Markov chain
   approximation for the
   mixed regular-singular control model with regime-switching. The
discrete-time and finite-state controlled Markov
chain is so defined that it is
locally consistent with
   \eqref{1d-state}.
Note that the state of the   process has two components $x$
and $\al$.  Hence in
order to use the methodology in \cite{Kushner-D},
 our  approximating Markov chain must have two components: one component
delineates the diffusive behavior
 whereas the other keeps track of the regimes.
Let $h>0$ be a discretization parameter. Define $L_h = \{ x: x = kh,
k = 0, \pm 1, \pm 2, \dots\}$ and $S_h= L_h \cap G_h$, where $G_h=(0, B+h)$ and
$B$ is an upper bound introduced for numerical computation purpose. Moreover,
assume
 without loss of generality that the boundary point   $B$
  is an integer multiple of $h$.
  Let $\{(\xi_n^h, \al_n^h), n<\infty
\}$ be a controlled discrete-time Markov chain on
 $S_h \times \M$ and   denote  by
$p^h((x,\ell),(y,\iota)|\pi^h)$ the transition probability from a
state $(x,\ell) $ to another state $(y,\iota) $ under the control
$\pi^h$. We need to define $p^h$ so that the chain's evolution
well approximates the local behavior of the controlled
regime-switching diffusion \eqref{1d-state}. At any discrete time
$n$, we can either exercise a regular control,    a singular
control or a reflection step. That is, if we put  $\Delta \xi_n^h
= \xi_{n+1}^h -\xi_n^h$, then \beq{Delta-xi-n}\Delta \xi_n^h
=\Delta \xi_n^hI_{\set{\text{regular control step at }n}} +
\Delta \xi_n^hI_{\set{\text{singular control step at }n}} + \Delta
\xi_n^hI_{\set{\text{reflection step at }n}} .\eeq The chain and
the control will be chosen so that there is exactly  one  term in
\eqref{Delta-xi-n} is nonzero. Denote by $\set{I_n^h:
n=0,1,\dots}$ a sequence of control actions, where $I_n^h=0, 1$ or
$2$, if  we exercise a  singular control, regular control, or
reflection at time  $ n$, respectively.

If  $I_n^h =1$,  then we  denote by $u_n^h\subset U$
the random variable that is the regular control action for the chain at
  time $n$.
Let $\tilde \Delta
t^h(\cdot,\cdot,\cdot)>0$ be the {\em interpolation interval} on $S_h \times \M
\times U$.
Assume
$\inf_{x,\ell,u} \tilde \Delta t^h(x,\ell,u)>0$ for each $h>0$ and
$\lim_{h\to 0} \sup_{x,\ell,u}\tilde \Delta t^h(x,\ell,u) \to 0$.

Let
$\ex_{x,\ell,n}^{u,h,1}$, ${\var}_{x,\ell,n}^{u,h,1}$ and
$\pr_{x,\ell,n}^{u,h,1}$ denote the conditional expectation, variance,
and marginal probability given $\{\xi_k^h, \al_k^h, u_k^h, I_k^h, k\le n,
\xi_n^h = x, \al_n^h = \ell,I_n^h=1, u_n^h = u\}$, respectively. The
sequence $\{(\xi_n^h,\al_n^h)\}$ is said to be \textit{locally
consistent},
if it satisfies

 \begin{equation*} \label{dfn-lc} \barray
\ad \ex_{x,\ell,n}^{u,h,1}  [\Delta \xi_n^h  ]= b(x,\ell,u)  \tilde \Delta
t^h(x,\ell,u) + o( \tilde\Delta t^h(x,\ell,u)), \\[1.5ex]
\ad \var_{x,\ell,n}^{u,h,1} (\Delta \xi_n^h )= \sg^2(x,\ell,u)  \tilde \Delta
t^h(x,\ell,u) + o( \tilde \Delta t^h(x,\ell,u)),
\\[1.5ex] \ad \pr_{x,\ell,n}^{u,h,1} \{\al_{n+1}^h = \iota\} = q_{\ell\iota}(x)
\tilde\Delta t^h(x,\ell,u)
 + o( \tilde\Delta t^h(x,\ell,u)), \text{ for }  \iota \neq \ell, \\[1.5ex]
\ad \pr_{x,\ell,n}^{u,h,1} \{\al_{n+1}^h = \ell\} =   (1+q_{\ell\ell}(x)) \tilde
\Delta
t^h(x,\ell,u) + o( \tilde\Delta
t^h(x,\ell,u)).
 \\[1.5ex]
 \ad \sup_{n, \omega \in \Omega} |\Delta \xi_{n}^h| \to 0 \ \hbox{ as
  } \ h\to 0.
 \earray
\end{equation*}

 If $I_n^h =0$,  then we  denote by $\Delta z_n^h $
the random variable that is the singular control action for the chain at
  time $n$ if $\xi_n^h \in [0, B]$.
  Note that $\Delta \xi_n^h = -\Delta z_n^h =
- h$.
If $I_n^h =2$, or $\xi_n^h= B+h$, reflection step is exerted definitely.
Dividend
is paid out to lower the surplus level. Moreover, we require reflection takes
the state from $B+h$ to $B$. That is, if we denote by $\Delta g_n^h $
the random variable that is the reflection action for the chain at
  time $n$, then  $\Delta \xi_n^h = -\Delta g_n^h =  -h$.

The singular control can be seen as a combination of
``inside" part ($I_n^h =0$) and ``boundary" part ($I_n^h =2$).
Also we require the singular control and reflection to be ``impulsive''
  or ``instantaneous.''
In other words, the interpolation interval on $S_h \times \M \times U\times
\set{0,1,2}$
 is   \beq{defn-interpolation-interval}\Delta
 t^h(x,\ell,u,i)= \tilde \Delta
t^h(x,\ell,u) I_{\set{i=1}},
  \text{ for any } (x,\ell,u,i) \in S_h \times \M \times U\times
\set{0,1,2}.\eeq
 Denote by $\pi^h:=\{\pi_n^h, n\ge 0 \}$ the sequence of control actions, where
  $$\pi_n^h:= \Delta z^h_n I_{\{I_n^h=0\}} + u_n^h I_{\{I_n^h=1\}}+  \Delta
g^h_n I_{\{I_n^h=2\}}.$$ The sequence $\pi^h$ is said to be {\em admissible}
  if $\pi_n^h$ is $\sigma\set{(\xi_0^h,\al_0^h),\dots,(\xi_n^h,\al_n^h),
\pi_0^h,\dots, \pi^h_{n-1}}$-adapted and
  for any $E\in \mathcal B(S_h\times\M)$, we have
  $$\pr\set{(\xi_{n+1}^h,\al_{n+1}^h) \in E
\big|\sigma\{(\xi_0^h,\al_0^h),\dots,(\xi_n^h,\al_n^h), \pi_0^h,\dots,
\pi^h_{n}\}}=
  p^h((\xi_n^h,\al_n^h), E|\pi_n^h), $$
  and $$ \pr\set{(\xi_{n+1}^h,
\al^h_{n+1})=(B,\ell)\big|(\xi_n^h,\al_n^h)=(B+h,\ell),\sigma\{(\xi_0^h,\al_0^h)
,\dots,(\xi_n^h,\al_n^h), \pi_0^h,\dots, \pi^h_{n}\} } =1.  $$ Put
$$t_0^h:=0, \ t_n^h: = \sum_{k=0}^{n-1} \Delta  t^h (\xi_k^h,
\al_k^h, u_k^h, I_k^h),\  \text{ and  }n^h(t):= \max\set{n: t_n^h \le t}.$$
Then
the piecewise constant interpolations,
denoted by $(\xi^h\cd, \al^h\cd)$, $u^h\cd$, $g^h\cd$, and $z^h\cd$, are
naturally defined as
\beq{def-interp-1}
\xi^h(t) = \xi_n^h, \  \al^h(t) = \al_n^h, \ u^h(t) = u_n^h, \ g^h(t) =
\sum_{k\le n^h(t)} \Delta g_k^h I_{\set{I_k^h=2}}, \
z^h(t) = \sum_{k\le n^h(t)} \Delta z_k^h I_{\set{I_k^h=0}},
\eeq for $t\in [t_n^h, t_{n+1}^h)$.
Let $\eta_h := \inf\set{n: \xi_n^h \in \partial G}$. Then the first exit time of
$\xi^h$ from $G$ is $\tau^h= t^h_{\eta_h}$. Let $(\xi_0^h,\al_0^h)=(x,\ell)\in
S_h\times \M$ and $\pi^h$ be an admissible control. The cost function for the
controlled  Markov chain is defined as
\beq{costdiscrete}
J_B^h(x,\ell,\pi^h) = \ex \sum_{k=1}^{\eta_h  -1 }e^{-r
t_k^h}[f( \xi_k^h, \al_k^h, u_k^h) \Delta t_k^h + c( \xi_k^h, \al_k^h)\Delta
z_k^h ],
\eeq
which is analogous to \eqref{payoff-numerical} thanks to the
definition of
 interpolation intervals in \eqref{defn-interpolation-interval}.
 The value function of the controlled Markov chain is
 \beq{value-discrete-chain}
 V^h_B(x,\ell)= \disp\sup_{\pi^h \text{ admissible}} J^h_B(x,\ell,\pi^h).
 \eeq
We shall show that  $V^h_B(x,\ell)$ satisfies the dynamic
programming equation:
 \beq{Vh-dynamic-eq}
V^h_B(x,\ell) = \left\{\barray \ad \disp\max_{u\in U}
\bigg\{\sum_{(y,\iota)} e^{-r \Delta t^h (x,\ell,u,
1)}p^h((x,\ell),(y,\iota)|\pi ) V^h(y,\iota) + f(x,\ell,u)
\Delta t^h(x,\ell,u,1) , \\  \aad \qquad \qquad  \big
[\disp\sum_{(y,\iota)} p^h((x,\ell),(y,\iota)|\pi) V^h(y,\iota)
+ c(x,\ell)h \big] \bigg\}, \ \    \text{ for } x \in S_h,\\
\aad 0,\ \hfill  \text{ for } x=0. \earray\right. \eeq Note that
discounting does not appear in the second line above because singular control is impulsive.
In the actual computing,
we use
iteration in value space or iteration in policy
space together
 with Gauss-Seidel
 iteration to solve $V^h$.
 The computations will be very involved. In contrast to the usual state space
$S_h$ in \cite{Kushner-D},  here we need to deal with an enlarged state space
$S_h \times \M$
  due to the presence of regime switching.

Define
the approximation to the first and the second derivatives of
$V(\cdot, \ell)$ by finite difference method in the first part of
QVIs \eqref{qvi} using stepsize $h>0$ as: \beq{finite-d} \barray
\ad V(x,\ell) \to   V^h(x,\ell)  \\
\ad V_x(x,\ell) \to \frac {V^h(x+h, \ell)-V^h(x, \ell)}
{h} \ \hbox{ for } \  b(x,\ell,u) >0 ,\\
\ad V_x(x,\ell) \to \frac {V^h(x,\ell)-V^h(x-h, \ell)} {h} \
\hbox{ for } \ b(x,\ell,u)<0, \\
\ad V_{xx}(x,\ell) \to \frac {V^h(x+h,\ell)-2V^h(x, \ell) + V^h(x-h,
\ell)} {h^2}. \earray\eeq
For the second part of the QVIs, we choose
$$ V_x(x,\ell) \to \frac {V^h(x,\ell)-V^h(x-h, \ell)} {h}. $$
Together with the boundary conditions, it
leads to \beq{finite-de} \barray
\ad V^h(x,\ell)=0, \ \hbox{for}\ x=0,  \\
\ad \max_{u\in U}\Big\{ \frac {V^h(x+h, \ell)-V^h(x, \ell)} {h} b(x,\ell,u)^+-
\frac
{V^h(x,\ell)-V^h(x-h, \ell)} {h}b(x,\ell,u)^-\\
\aad \ \ +\frac {V^h(x+h,\ell)-2V^h(x, \ell) + V^h(x-h, \ell)}
{h^2} \frac{\sg^2(x,\ell,u)}{2}\\
\aad \ \ +\sum_\iota V^h(x,\cdot)q_{\ell\iota}-r
V^h(x,\ell)+u,\ \ \  c(x,\ell)- \frac {V^h(x,\ell)-V^h(x-h, \ell)}
{h}\Big\}=0,\\
\aad \qquad \qquad \forall x\in G_h^o, \ell\in\cal{M}, \earray\eeq
where $b(x,\ell,u)^+$ and $b(x,\ell,u)^-$ are the positive and
negative parts of $b(x,\ell,u)$, respectively. Simplifying
\eqref{finite-de} and comparing the result with
\eqref{Vh-dynamic-eq}, we achieve the transition probabilities of
the first part of the right side of \eqref{Vh-dynamic-eq} as the
following: \beq{trans}\barray \ad p^h((x,\ell),(x+h,\ell)|\pi)=
\frac{(\sg^2(x,\ell,u)/2)+hb(x,\ell,u)^+}{D
-r h^2},\\
\ad p^h((x,\ell),(x-h,\ell)|\pi)=\frac{(\sg^2(x,\ell,u)/2)+hb(x,\ell,u)^-
}{D-r h^2},\\
\ad p^h((x,\ell),(x,\iota)|\pi)=\frac{h^2}{D-r
h^2}q_{\ell\iota},\ \
\hbox{for}\ \ \ell\neq\iota,\\
\ad p^h\cd=0,\ \ \hbox{otherwise},\\
\ad \Delta t^h(x,\ell,u, 1)=\frac{h^2}{D}, \earray\eeq
\hbox{with} $$
D=\sg^2(x,\ell,u)+h|b(x,\ell,u)|+h^2(r-q_{\ell\ell})$$ being
well defined. We also find the transition probability for the
second part of the right side of \eqref{Vh-dynamic-eq}.
That is,
$$p^h((x,\ell),(x-h,\ell)|\pi)=1.$$

\rem The transition probabilities are quite natural. The first part of the QVIs
can be seen as a ``diffusion" region, where the regular control is dominant.
The Markov approximating chain can switch between regimes and states nearby.
But the second part of the QVIs is the ``jump" region, where the dividends are
paid out and the singular control is dominant. The singular control will
project the Markov approximation chain back one step $h$ w.p.1 due to the
representation.

Since the wealth can not reach infinity, we need only need choose
$B$ large enough and compute the value function in the finite
interval $[0, B]$. Our ultimate goal is to show $V^h$ converges to
$V$ in a large enough interval $[0, B]$ as $h \to 0$. A common
approach (\cite{Kushner-D}) is to show that the collection
$\set{(\xi^h,\al^h), u^h, g^h, z^h, h\ge 0}$ is tight and then
appropriately characterize the subsequential weak limit. However,
the above scheme is problematic since in general, the processes
$\set{g^h\cd, z^h\cd, h\ge 0}$ may fail to be tight. To overcome
this difficulty, we adapt the techniques developed in
\cite{Kushner-M-1991} and \cite{Budhiraja-Ross}. The basic idea is
to (a) suitably re-scale the time so that the  processes involved
in the convergence analysis are tight in the new time scale; (b)
carry out  weak convergence analysis with the rescaled processes;
and (c) revert back to the original time scale to obtain the
convergence of $V^h $ to $V$. Note that the setting in our problem
is different from those in the aforementioned references.
Moreover, the presence of regime switching adds additional
difficulty in the analysis.

\section{Convergence of Numerical Approximation}\label{sec:conv}

This section focuses on the asymptotic properties of the
approximating Markov chain proposed in the last section. The main
techniques are methods of weak convergence. To begin with, the
technique of time rescaling is given in Section \ref{sec:41}. The
interpolation of the approximation sequences
is  introduced in Section \ref{sec:41}. The
definition of relax controls and chattering lemmas of optimal
control are presented in Sections \ref{sec:42} and \ref{sec:43},
respectively. Section \ref{sec:44} deals with weak convergence of
 $\{\hat \xi^h\cd, \hat \al^h\cd,
\hat m^h\cd, \hat w^h\cd, \hat z^h\cd, \hat g^h\cd, \hat
T^h\cd\}$, a sequence of rescaled process. As a result, a sequence
of controlled surplus processes converges to a limit surplus
process. By using the techniques of inversion, Section
\ref{sec:44} also takes up the issue of the weak convergence of
the surplus
process. Finally Section \ref{sec:45} 
establishes the convergence of the value function.

\subsection{Interpolation and Rescaling}\label{sec:41}

Based on the approximating Markov chain constructed above, the
piecewise constant interpolation is obtained and the appropriate interpolation
interval
level is chosen. Recalling \eqref{def-interp-1}, the continuous-time
interpolations
$(\xi^h\cd, \al^h\cd)$, $u^h\cd$, $g^h\cd$, and $z^h\cd$ are defined. In
addition,
let
${\cal{U}}^h$ denote the collection of controls, which are
determined by a sequence of measurable functions $F_n^h(\cdot)$
such that \beq{cmf} u_n^h=F_n^h(\xi_k^h,  \alpha_k^h,k\leq
n;u_k^h,k\le n).\eeq
Define ${\cal {D}}_t^h$ as the smallest
$\sigma$-algebra generated by
$\{\xi^h(s),\alpha^h(s), u^h(s), g^h(s), z^h(s),$
 $s\leq t\}.$ In addition, $\cal{U}$$^h$ defined by \eqref{cmf} is
equivalent to the collection of all piecewise constant admissible
controls with respect to $\cal{D}$$_{t}^h.$

Using the representations of regular control, singular control,
reflection step and the interpolations defined above, \eqref{Delta-xi-n}
yields

\beq{lcl} \barray \xi^h(t) \ad
=x+\sum_{k=0}^{n-1}[E_k^h\Delta\xi_k^h+(\Delta\xi_k^h-
E_k^h\Delta\xi_k^h)]-z^h(t)-g^h(t)\\
\ad =x+\sum_{k=0}^{n-
1}b(\xi_k^h, \alpha_k^h, u_k^h)\Delta
t^h(\xi_k^h,\alpha_k^h,u_k^h) +\sum_{k=0}^{n-1}(\Delta\xi_k^h-
E_k^h\Delta\xi_k^h)-z^h(t)-g^h(t)+\varepsilon^h(t)\\
\ad
=x+ B^h(t)+ M^h(t)-z^h(t)-g^h(t)+\varepsilon^h(t),
\earray\eeq where
$$B^h(t)=\sum_{k=0}^{n-1}b(\xi_k^h, \alpha_k^h, u_k^h)\Delta
t^h(\xi_k^h,\alpha_k^h,u_k^h),$$
$$M^h(t)=\sum_{k=0}^{n-1}(\Delta\xi_k^h-
E_k^h\Delta\xi_k^h), $$ and $\varepsilon^h(t)$ is a negligible error
satisfying \beq{error-eh} \lim_{h\to\infty}\sup_{0\leq t\leq
T}E|\varepsilon^h(t)|^2\to0 \ \hbox{ for any }
 \ 0<T<\infty.\eeq
Also, $M^h(t)$ is a martingale with respect to $\cal{D}$$_t^h$,
and its {\em discontinuity goes to zero} as $h\to0.$ We attempt to represent $M^h(t)$ in
a form similar to the diffusion term in \eqref{1d-state}. Define
$w^h(\cdot)$ as \beq{diff} \barray w^h(t)\ad
=\sum_{k=0}^{n-1}(\Delta\xi_k^h-
E_k^h\Delta\xi_k^h)/\sigma(\xi_k^h,\al_k^h, u_k^h),\\
\ad =\int_0^t \sigma^{-1}(\xi^h(s),\al^h(s),u^h(s))dM^h(s).\earray \eeq
We can now rewrite \eqref{lcl} as \beq{wlx}
\xi^h(t)=x+\int_0^tb(\xi^h(s), \al^h(s),
u^h(s))ds+\int_0^t\sigma(\xi^h(s),\al^h(s), u^h(s))
dw^h(s)-z^h(t)-g^h(t)+\varepsilon^h(t).\eeq

Next we will introduce the rescaling process. The basic idea of rescaling time is
to ``stretch out" the control and state processes so that they are ``smoother''
and therefore the tightness of $g^h\cd$ and $z^h\cd$ can be proved. Define $\Delta \hat
t_n^h$
by
 \beq{rescale}
  \Delta \hat t_n^h=\begin{cases}
\Delta t^h \ \ \ \hbox{ for a diffusion on step $n$},\\
|\Delta z_n^h|=h \ \ \ \hbox{ for a singular control on step $n$}, \\
|\Delta g_n^h|=h \ \ \ \hbox{ for a reflection on step $n$},
 \end{cases}
\eeq
Define $\hat T^h\cd$ by
$$\hat T^h(t)=\sum_{i=0}^{n-1}\Delta t^h=t_n^h, \ \ \ \hbox{for} \ \ t\in[\hat
t_n^h, \hat t_{n+1}^h]$$
Thus, $\hat T^h\cd$ will increase with the slope of unity if an only if a regular
control is exerted.
In addition, define the rescaled and interpolated process
$\hat \xi^h(t)=\xi^h(\hat T^h(t))$, likewise define
$\hat \al^h(t))$, $\hat u^h(t)$, $\hat g^h(t)$ similarly.
The time scale is stretched out by $h$ at the
reflection and singular control
steps. We can now write
\beq{wlxhat}\begin{aligned}
\hat \xi^h(t)=x & +\int_0^tb(\hat \xi^h(s), \hat \alpha^h(s), \hat
u^h(s))ds\\ & +\int_0^t\sigma( \hat \xi^h(s),\hat \al^h(s), \hat u^h(s))
dw^h(s) -\hat z^h(t)-\hat g^h(t)+\varepsilon^h(t).
\end{aligned}\eeq

\subsection{Relaxed Controls}\label{sec:42}
Let ${\cal B}({U} \times[0,\infty))$ be the $\sigma$-algebra of
Borel subsets of ${U} \times [0,\infty)$. An {\it admissible relaxed
control} (or deterministic relaxed control) $m(\cdot)$ is a measure
on ${\cal B}({U} \times[0,\infty))$ such that $m({U}\times[0,t]) =
t$ for each $t \ge 0$. Given a relaxed control $m(\cdot)$, there is
an $m_t(\cdot)$ such that $m(d\phi dt) = m_t(d\phi )dt$. We can
define $m_t(B) = \lim_{\delta \to 0} {\frac {m(B\times[t-\delta,t])}
{\delta}}$ for $B\in {\cal B}(U)$. With the given probability space,
we say that $m\cd$ is an admissible relaxed (stochastic) control for
$(w\cd, \al\cd)$ or $(m\cd,w\cd,\al\cd)$ is admissible, if
$m(\cdot,\omega)$ is a deterministic relaxed control with
probability one and if $m(A\times [0,t])$ is ${\cal F}_t$-adapted
for all $A \in {\cal B}(U)$. There is a derivative $m_t\cd$ such
that $m_t\cd$ is ${\cal F}_t$-adapted for all $A \in {\cal B}(U)$.

 Given a relaxed control $m(\cdot)$ of $u^h(\cdot)$, we define
the derivative $m_t\cd$ such that \beq{def-eq} m^h(K)= \int_{U
\times [0,\infty)} I_{\{(u^h, t) \in K\}}m_t(d\phi ) dt \eeq for all
 $K
\in {\cal B}(U\times [0,\infty))$, and that for each $t,$ $m_t\cd$
is a measure on ${\cal B}(U)$ satisfying $m_t(U)=1$. For example, we
can define $m_t\cd$ in any convenient way for $t=0$ and as the
left-hand derivative for $t>0$, \beq{mtA} m_t(A) = \lim_{\delta \to
0} \frac {m(A\times[t-\delta,t])} {\delta}, \ \forall A\in {\mathcal
B}(U).\eeq  Note that $m(d\phi  dt) = m_t(d\phi )dt$. It is natural
to define the relaxed control representation $m^h(\cdot)$ of
$u^h(\cdot)$ by \beq{mth} m_t^h(A)=I_{\{u^h(t)\in A\}}, \ \forall
A\in {\mathcal B}(U). \eeq

Let ${\cal F}_t^h$ be  a filtration, which denotes the minimal
$\sigma$-algebra that measures \beq{sga}
\{\xi^h(s),\alpha^h(\cdot),m_s^h(\cdot),w^h(s),z^h(s),g^h(s), s\leq
t\}.\eeq Use ${\Gamma}^h$ to denote the set of admissible relaxed
controls $m^h(\cdot)$ with respect to
$(\alpha^h(\cdot),w^h(\cdot))$ such that $m_t^h(\cdot)$ is a
fixed probability measure in the interval $[t_n^h, t_{n+1}^h)$ given
${\cal F}_t^h$. Then $\Gamma^h$ is a larger control space containing
${\cal U}^h$.
Referring to the stretched out time scale, we denote the rescaled
relax control as $m_{\hat T^h(t)}(d\phi)$. Define $M_t(A)$ and $M^h_t(d\phi)$
by
$$ \barray \ad M_t(A)dt=dw(t)I_{u(t)\in A}, \ \forall
A\in {\mathcal B}(U) \\
\ad  M^h_t(d\phi)dt=dw^h(t)I_{u^h(t)\in \cal U}. \earray $$ Analogously, as an
extension of time rescaling, we let
$$\hat M^h_{\hat T^h(t)}(d\phi)d\hat T^h(t)=d \hat w^h(\hat T^h(t))I_{u^h(\hat
T^h(t))\in \cal U}.$$ With the notation of relaxed control given
above, we can write \eqref{wlx}, \eqref{wlxhat} and the value
function \eqref{value-numerical-purpose} as \beq{wlxrelax}
\xi^h(t)=x+\int_0^t\int_{\cal U}b(\xi^h(s), \al^h(s),
\phi)m_s^h(d\phi )ds+ \int_0^t\int_{\cal
U}\sigma(\xi^h(s),\al^h(s), \phi)M^h_s(d\phi)ds+\varepsilon^h(t),
\eeq
\beq{wlxhatrelax}\barray  \hat
\xi^h(t)=\ad x+\int_0^t\int_{\cal U}b(\hat \xi^h(s), \hat \al^h(s),
\phi)\hat m_{\hat T^h(s)}^h(d\phi
)d\hat T^h(s) \\
\aad \hfill +\int_0^t\int_{\cal U}\sigma( \hat \xi^h(s),\hat \al^h(s), \phi)
\hat M_{\hat T^h(s)}(d\phi)d\hat T^h(s) -\hat z^h(t)-\hat
g^h(t)+\varepsilon^h(t), \earray \eeq
and \beq{revalue}
V^h(x,\ell)=\inf_{m^h\in \Gamma^h}J^h(x,\ell,m^h). \eeq Now
we  give the definition of existence and uniqueness of weak
solution.
\begin{defn}\label{relcon}
   {\rm By a weak solution of \eqref{wlxrelax}, we mean that
   there exists a probability space $(\Omega,{\cal F},P)$, a
   filtration ${\cal F}_t$, 
   and process
   $(x\cd,\al\cd,m\cd,w\cd)$ such that
   $w\cd$ is a standard ${\cal F}_t$-Wiener process,
   $\al\cd$ is a Markov chain with generator $Q$ and state
   space ${\cal M}$, $m\cd$ is admissible with respect to
   $x\cd$, and is ${\cal F}_t$-adapted, and \eqref{wlxrelax} is satisfied. For an
   initial condition $(x,\ell)$, by the
   weak sense uniqueness, we mean
   that the probability law of the admissible
   process $(\alpha\cd, m\cd,w\cd)$ determines the
   probability law of solution $(x\cd,\alpha\cd,
   m\cd,w\cd)$ to \eqref{wlxrelax}, irrespective of
   probability space.
} \end{defn}

To proceed, we need some assumptions.

\begin{itemize}
\item[(A1)] Let $u\cd$ be an admissible ordinary control with
respect to $w\cd$ and $\alpha\cd$, and suppose that
$u\cd$ is piecewise constant and takes only a finite number of
values. For each initial condition, there exists a solution to
\eqref{wlxrelax} where $m\cd$ is the relaxed control
representation of $u\cd$. This solution is unique in the weak
sense.
\end{itemize}

\subsection{A Chattering Lemma and Approximation to the Optimal
Control}\label{sec:43}

 This section deals with the approximation of relaxed
controls by ordinary controls. We can always use relax controls
to approximate the ordinary controls, which is only a tool for
mathematical analysis. Here we present a result
of chattering lemma for our problem. The proof of the chattering
lemma can be found in \cite{Kushner90}.

\begin{prop}\label{chattering}
Let $(m\cd,w\cd)$ be admissible for the problem given in
{\rm(\ref{wlxrelax})}. Then given $\varsigma>0$, there is a finite set
$\{\gamma^\varsigma_1,\ldots,\gamma^\varsigma_{l_\varsigma}\}=U^\varsigma
\subset U$,
and an $\e  >0$ such that there is a probability space on which are
defined $(x^\varsigma\cd,\alpha^\varsigma\cd,
u^\varsigma\cd,w^\varsigma\cd)$,
where
$w^\varsigma\cd$ are standard Brownian motions, and $u^\varsigma\cd$ is
an admissible $U^\varsigma$-valued ordinary control on the interval $
[k\e ,k\e +\e )$. Moreover, \beq{est-cha}\barray \ad P^m_x \(\sup_{
s\le T} |x^\varsigma(s)- x(s)| > \varsigma \) \le
\varsigma, \ and \\
\ad |J^m_x \cd- J^{u^\varsigma}_x\cd| \le \varsigma.\earray\eeq
\end{prop}

Coming back to the approximation to the optimal control, to show that the
discrete approximation of the value function $V^{h}(x,\ell)$
converges to the value function $V(x,\ell)$, we shall use the
comparison control techniques. In doing so, we need to verify
certain continuity properties. The details of the proof is presented
in the appendix.

\begin{prop}\label{approx}
For {\rm(\ref{wlxrelax})}, let $\wdt \varsigma >0$ be
given and $(x\cd,\alpha\cd, m\cd,w\cd)$ be an $\wdt
\varsigma$-optimal
control. For each $\varsigma>0$, there is an $\e>0$ and a probability
space on which are defined $w^\varsigma\cd$, a control $u^\varsigma\cd$
as in \thmref{chattering}, and a solution $x^\varsigma\cd$ such that the
following assertions hold:
\begin{itemize}
\item[{\rm(i)}]
\beq{diff-www} |J^m_x \cd- J^{u^\varsigma}_x\cd|\le \varsigma.\eeq
\item[{\rm (ii)}]
Moreover, there is a $\theta>0$ such that the approximating $u^\varsigma\cd$
can be chosen so that its probability law at $n\e$, conditioned on
$\{w^\varsigma(\tau), \alpha^\varsigma(\tau), \tau \le n\e; u^\varsigma (k \e),
k < n\}$
depends
only on the samples $\{w^\varsigma(p \theta), \alpha^\varsigma(p\theta), p
\theta \le n\e;
u^\varsigma(
k\e), k < n\}$, and is continuous in the $w^\e(p\theta)$
arguments.
\end{itemize}
\end{prop}

\subsection{Convergence of A Sequence of Surplus Processes}\label{sec:44}

\begin{lem}\label{conse1}
Using the transition probabilities $\{p^h\cd\}$ defined in
{\rm\eqref{trans}}, the interpolated process of the constructed
Markov chain $\{\hat \al^h\cd\}$ converges weakly to
$\hat \al\cd$, the Markov chain with generator
$Q=(q_{\ell\iota})$. \end{lem}

\para{Proof.} It can be seen that $\al^h\cd$
is tight. The proof can be obtained
similar to
Theorem 3.1 in \cite{YinZB}. That is,
\beq{tightness1}
\l\ex [(\al^h(t+s)-\al^h(t))^2]|{\cal F}_t^h\r \leq \tilde\gamma(s)
\ \ \hbox{and} \ \  \lim_{s \to 0}\lim_{h \to 0}\sup\ex \tilde\gamma(s)=0,
\eeq
where $\tilde\gamma(s)\geq 0$ is ${\cal F}_t^h$-measurable.
On the other hand, due to the definition of $\hat \al^h\cd$, we have
\beq{tightness2}
\l\ex [(\hat \al^h(t+s)-\hat \al^h(t))^2]|{\cal F}_t^h\r
\leq \l\ex [(\al^h(t+s)-\al^h(t))^2]|{\cal F}_t^h\r \leq \tilde\gamma(s).
\eeq
Combining \eqref{tightness1} and \eqref{tightness2}, we obtain $\hat \al^h \cd$ is tight.
Thus, the constructed
Markov chain $\{\hat \al^h\cd\}$ converges weakly to
$\hat \al\cd$. \qed

\begin{thm}\label{conse2}
 Let the approximating chain
$\{\xi_n^h,\al_n^h,n<\infty\}$ constructed with transition
probabilities defined in {\rm\eqref{trans}} be locally consistent
with {\rm\eqref{1d-state}}, $m^h(\cdot)$
be the relaxed control representation
of
$\{u_n^h,n<\infty\}$, $(\xi^h\cd,\al^h\cd)$ be the
continuous-time interpolation defined in {\rm\eqref{def-interp-1}},
and $\{\hat \xi^h\cd, \hat \al^h\cd, \hat m^h\cd,
\hat w^h\cd, \hat z^h\cd, \hat g^h\cd, \hat T^h\cd\}$ be the
corresponding rescaled processes.
Then $\{\hat \xi^h\cd, \hat \al^h\cd, \hat m^h\cd,
\hat w^h\cd, \hat z^h\cd, \hat g^h\cd, \hat T^h\cd\}$ is
tight.
\end{thm}

\para{Proof.}
In view of \lemref{conse1}, $\{\hat \al^h\cd\}$ is tight. The
sequence $\{\hat m^h\cd\}$ is tight since its range space is
compact.
 Let
$T<\infty$, and let $\tau_h$ be an ${\cal F}_t$-stopping time
which is no bigger than $T$. Then for $\dl>0$, \beq{wienercon}
E_{\tau_h}^{u^h}(w^h(\tau_h+\dl)-w^h(\tau_h))^2
=\dl+{\e}_h,\eeq where ${\e}_h\to0$ uniformly in
$\tau_h$. Taking $\limsup_{h\to0}$ followed by
$\lim_{\delta\to0}$ yield the tightness of $\{w^h\cd\}$. Similar to the
argument of $\al^h\cd$, the tightness of $\hat w^h\cd$ is obtained.
Furthermore, following
the definition of ``stretched out" timescale,
$$ \barray \ad
 |\hat z^h(\tau_h+ \delta)-\hat z^h(\tau_h)|\leq |\delta| + O(h),\\
\ad |\hat g^h(\tau_h+ \delta)-\hat g^h( \tau_h)| \leq |\delta|+
O(h). \earray $$ Thus $\{\hat z^h\cd, \hat g^h\cd\}$ is tight. For
notational simplicity, we assume that $b\cd$ and $\sigma\cd$ are
bounded. For more general case, we can use a truncation device.
These results and the boundedness of $b\cd$ implies the
tightness of $\{\xi^h\cd\}$. Therefore it follows that  $$\{\hat \xi^h\cd, \hat
\al^h\cd, \hat u^h\cd, \hat w^h\cd, \hat z^h\cd, \hat g^h\cd, \hat
T^h\cd\}$$ is tight.
 \qed

Since $\{\hat x^h\cd, \hat \al^h\cd, \hat m^h\cd,
\hat w^h\cd, \hat z^h\cd, \hat g^h\cd,
\hat T^h\cd\}$ is tight, we can extract a
weakly convergent
subsequence denoted by $\{\hat \xi\cd, \hat \al\cd, \hat m\cd,
\hat w\cd, \hat z\cd, \hat g\cd, \hat T\cd\}$. Also,
the paths of $\{\hat x\cd, \hat \al\cd, \hat m\cd,
\hat w\cd, \hat z\cd, \hat g\cd, \hat T\cd\}$ are
continuous w.p.1.

\begin{thm}\label{conse3}
Let $\{\hat x\cd, \hat \al\cd, \hat m\cd,
\hat w\cd, \hat z\cd, \hat g\cd, \hat T\cd\}$ be the limit of
weakly convergent subsequence of $\{\hat \xi^h\cd, \hat \al^h\cd, \hat m^h\cd,
\hat w^h\cd, \hat z^h\cd, \hat g^h\cd, \hat T^h\cd\}$. $w\cd$ is a
standard ${\cal F}_t$-Wiener process, and $m\cd$ is admissible.
Let $\hat {\cal F}_t$ the
$\sigma$-algebra generated by $\{\hat \xi^h\cd, \hat \al^h\cd, \hat m^h\cd,
\hat w^h\cd ,$ $\hat z^h\cd, \hat g^h\cd, \hat T^h\cd\}$. Then $\hat w(t)=w(\hat
T(t))$ is an
$\hat {\cal F}_t$-martingale with quadratic variation $\hat T(t)$.
The limit processes satisfy
\beq{wlxhatconv}
\barray
\hat x(t)\ad =x+\int_0^t\int_{\cal U}b(\hat x(s), \hat \al(s), \phi)\hat m_{\hat
T(s)}^h(d\phi
)d\hat T(s) \\
\aad \ +\int_0^t\int_{\cal U}\sigma( \hat x(s),\hat \al(s), \phi)
\hat M_{\hat T(s)}(d\phi)d\hat T(s) -\hat z(t)-\hat g(t). \earray
\eeq
\end{thm}

\para{Proof.}
For $\delta>0$, define the process $l(\cdot)$ by
$l^{h,\delta}(t)=l^h(n\delta), t\in[n\delta, (n+1)\delta)$. Then, by
the tightness of $\{\hat \xi^h(\cdot), \hat \al^h(\cdot)\}$,
\eqref{wlxhatrelax}
can be rewritten as \beq{rewlxhat}\barray
\hat \xi^h(t)\ad=x+\int_0^t\int_{\cal U}b(\hat \xi^h(s), \hat \al^h(s), \phi)\hat
m_{\hat T^h(s)}^h(d\phi
)d\hat T^h(s) \\
\aad \quad +\int_0^t\int_{\cal U}\sigma( \hat \xi^{h,\delta}(s),\hat
\al^{h,\delta}(s), \phi)
\hat M_{\hat T^h(s)}(d\phi)d\hat T^h(s) -\hat z^h(t)-\hat g^h(t)+\varepsilon^{
h,\delta}(t), \earray \eeq where
\beq{3}\lim_{\delta\to0}\limsup_{h\to0}E|\varepsilon^{h,\delta}(t)|=0.
\eeq

If we can verify $\hat w\cd$ to be an $\hat {\cal F}_t$-martingale, then
\eqref{wlxhatconv}
could be obtained by taking limits in \eqref{rewlxhat}.
To characterize $w(\cdot)$, let $t>0,$ $\delta>0$, $p$, $q$,
$\{t_k:k\leq p\}$ be given such that $t_k\leq t\leq t+s$ for all
$k\leq p$, $\psi_j(\cdot)$ for $j\leq q$ is real-valued and continuous
functions on $U\times[0,\infty)$ having compact support for all
$j\leq q$. Define \beq{defg} (\psi_j, \hat m)_t=\int_0^t\int_{\cal U}
\psi_j(\phi,s)\hat m_{\hat T(s)}^h(d\phi
)d\hat T(s). \eeq Let $S(\cdot)$ be a real-valued and
continuous function of its arguments with compact support. By
\eqref{diff}, $w^h(\cdot)$ is an ${\cal F}_t$-martingale. In view of
the definition of $\hat w(t)$,
we have \beq{1}
ES(\hat \xi^h(t_k),\hat \al^h(t_k), \hat w^h(t_k),(\psi_j,m^h)_{t_k},
\hat z^h(t_k),  \hat g^h(t_k), j\leq q,
k\leq p) [\hat w^h(t+s)- \hat w^h(t)]=0. \eeq By using the Skorohod
representation and the dominant convergence theorem, letting
$h\to0,$ we obtain \beq{2a}
ES(\hat \xi^h(t_k),\hat \al^h(t_k), \hat w^h(t_k),(\psi_j,m^h)_{t_k},
\hat z^h(t_k),  \hat g^h(t_k), j\leq q,
k\leq p) [\hat w(t+s)- \hat w(t)]=0. \eeq Since $\hat w(\cdot)$ has
continuous sample paths, \eqref{2a} implies that $\hat w(\cdot)$ is
a continuous ${\cal F}_t$-martingale. On the other hand, since
\beq{wienercon1} E[((\hat w^h(t+\delta))^2-(\hat w^h(t))^2]
=E[(\hat w^h(t+\delta)-\hat w^h(t))^2]
=\hat T(t+s)- \hat T(t), \eeq by using the Skorohod representation and the
dominant convergence theorem together with \eqref{wienercon1}, we
have \beq{2} \barray \ad ES(\hat \xi^h(t_k),\hat \al^h(t_k), \hat
w^h(t_k),(\psi_j,m^h)_{t_k},
\hat z^h(t_k),  \hat g^h(t_k), j\leq q,
k\leq p)\\ \aad \hfill \times
[\hat w^2(t+\delta)-\hat w^2(t)-(\hat T(t+s)- \hat
T(t))]=0. \earray \eeq The
quadratic variation of the martingale $\hat w(t)$ is $\Delta \hat T$, then
$\hat w(\cdot)$ is an $\hat {\cal F}_t$-Wiener process.

Let $h\to0$, by using the Skorohod representation, we obtain
\beq{con1}\barray \ad
E\left|\int_0^t\int_{\cal U}b(\hat \xi^h(s),
\hat \al^h(s), \phi)\hat m_{\hat
T^h(s)}^h(d\phi
)d\hat T^h(s) -
\int_0^t\int_{\cal U}b(\hat x(s),
\hat \al(s), \phi)\hat m_{\hat T(s)}^h(d\phi
)d\hat T(s)\right|=0 \earray \eeq uniformly in $t$ with probability one. On the
other hand, $\{\hat m^h(\cdot)\}$ converges in the compact weak topology,
that is, for any bounded and continuous function $\psi(\cdot)$ with
compact support,
as $h\to 0$, \beq{con2} \int_0^{\infty}\int_{\cal U}\psi(\phi,s)\hat m_{\hat
T^h(s)}^h(d\phi
)d\hat T^h(s)
\to \int_0^{\infty}\int_{\cal U}\psi(\phi,s)\hat m_{\hat T(s)}(d\phi
)d\hat T(s).  \eeq Again, the
Skorohod representation
(with  a slight abuse of notation)
implies that as $h\to 0$,
\beq{con3}\barray\ad
\int_0^t\int_{\cal U}b(\hat \xi^h(s), \hat \al^h(s), \phi)\hat m_{\hat
T^h(s)}^h(d\phi
)d\hat T^h(s)
\to \int_0^t\int_{\cal U}b(\hat x(s), \hat \al(s), \phi)\hat m_{\hat T(s)}(d\phi
)d\hat T(s)
 \earray \eeq uniformly in $t$ with probability one on any
bounded interval.

In view of \eqref{rewlxhat}, since $\xi^{h,\delta}(\cdot)$ and $\alpha^{h,\delta}(\cdot)$ are
piecewise constant functions,
\beq{con4}
\int_0^t\int_{\cal U}\sigma( \hat \xi^{h,\delta}(s),\hat \al^{h,\delta}(s), \phi)
\hat M_{\hat T^h(s)}(d\phi)d\hat T^h(s)
\to \int_0^t\int_{\cal U}\sigma( \hat x^\delta(s),\hat \al^\delta(s), \phi)
\hat M_{\hat T(s)}(d\phi)d\hat T(s) \eeq
as $h\to 0$. Combining
\eqref{defg}-\eqref{con4}, we have
\beq{xcon}\barray
\hat x(t)\ad=x+\int_0^t\int_{\cal U}b(\hat x(s), \hat \al(s), \phi)\hat m_{\hat
T(s)}^h(d\phi
)d\hat T(s) \\
\aad \quad +\int_0^t\int_{\cal U}\sigma( \hat x^\delta(s),\hat \al^\delta(s),
\phi)
\hat M_{\hat T(s)}(d\phi)d\hat T(s) -\hat z(t)-\hat
g(t)+\varepsilon^{\delta}(t),\earray \eeq where
$\lim_{\delta\to0}E|\varepsilon^{\delta}(t)|=0.$
Finally, taking limits in the above equation as $\delta\to0$,
\eqref{wlxhatconv} is obtained.  \qed

\begin{thm}\label{inverse}

For $t<\infty$, define the inverse
$$
R(t)=\inf\{s:\hat T(s)>t \}.
$$
Then $R(t)$ is right continuous and $R(t) \to \infty$ as $t \to \infty$ w.p.1.
For any process $\hat \varphi\cd$, define the rescaled process $\varphi\cd$ by
$\varphi(t)=\hat \varphi(T(t))$. Then, $w\cd$ is a standard ${\cal F}_t$-Wiener
process and \eqref{1d-state} holds.
\end{thm}

\para{Proof.}
Since $\hat T(t) \to \infty$ w.p.1 as $t \to \infty$,
$R(t)$ exists for all $t$ and $R(t) \to \infty $ as $t \to \infty$ w.p.1.
Similar
to \eqref{2a} and \eqref{2},
$$
ES( \xi^h(t_k), \al^h(t_k),  w^h(t_k),(\psi_j,m^h)_{t_k},
 z^h(t_k),   g^h(t_k), j\leq q,
k\leq p)\times[ w(t+s)-  w(t)]=0.
$$
$$
\barray \ad ES( \xi^h(t_k), \al^h(t_k),  w^h(t_k),(\psi_j,m^h)_{t_k},
 z^h(t_k),   g^h(t_k), j\leq q,
k\leq p)\\ \aad \hfill\times[ w^2(t+\delta)- w^2(t)-( R(t+s)-  R(t))]=0. \earray
$$
Thus, we can verify $w\cd$ is an ${\cal F}_t$-Wiener process.
A rescaling of \eqref{wlxhatconv} yields
\beq{wlxhatinverse}
\barray
 x(t)=x \ad +\int_0^t\int_{\cal U}b( x(s),  \al(s), \phi)m_s(d\phi
)ds \\ \ad +\int_0^t\int_{\cal U}\sigma(  x(s),\al(s), \phi)M_s(d\phi
)ds -z(t)- g(t). \earray
\eeq
In other words, \eqref{1d-state} holds. \qed

\subsection{Convergence of Cost and Value Functions}\label{sec:45}

\begin{thm}\label{costthem}
Let $h$ index the weak convergent subsequence of $\{\hat \xi^h\cd, \hat \al^h\cd,
\hat m^h\cd,
\hat w^h\cd, \\ \hat z^h\cd,  \hat g^h\cd, \hat T^h\cd\}$ with the limit
$\{\hat x\cd, \hat \al\cd, \hat m\cd,
\hat w\cd, \hat z\cd, \hat g\cd, \hat T\cd\}$. Then,
\beq{cost}
\barray
J^h(x,\ell,\pi^h)\ad  \to E_{x,\ell}^{\pi}\int_0^\tau\int_{\cal U}e^{-r\hat
T(t)}[f(\hat x(t), \hat \al(t), \phi)\hat m_t(d\phi)dt
+c(\hat x(t^-),\hat \al(t^-))d\hat Z] \\
\ad =E_{x,\ell}^{\pi}\int_0^\tau\int_{\cal U}e^{-rt}[f( x(t), \al(t), \phi)
m_t(d\phi)dt
+c(x(t^-),\al(t^-))dZ] = J(x,\ell, \pi).
\earray
\eeq
\end{thm}

\para{Proof.}
Note that $\Delta z^h = \Delta g^h=h$, the uniform integrability of $dZ$ can be
easily verified. Due to
the tightness and the uniform integrability properties, for any $t$,
$$\int_0^t c(\hat x(t^-),\hat \al(t^-))d\hat Z$$
can be well approximated by a Reimann sum uniformly in $h$.
By the weak convergence and the Skorohod representation,
$$
\barray J_B^h(x,\ell,\pi^h) \ad = \ex \sum_{k=1}^{\eta_h  -1 }e^{-r
t_k^h}[f( \xi_k^h, \al_k^h, u_k^h) \Delta t_k^h + c( \xi_k^h, \al_k^h)\Delta
z_k^h] \\ \ad
\to E_{x,\ell}^{\pi}\int_0^\tau\int_{\cal U}e^{-r\hat T(t)}[f(\hat x(t), \hat
\al(t), \phi)\hat m_t(d\phi)dt
+c(\hat x(t^-),\hat \al(t^-))d\hat Z]
.\earray
$$
By an inverse transformation,
$$
\barray \ad
E_{x,\ell}^{\pi}\int_0^\tau\int_{\cal U}e^{-r\hat T(t)}[f(\hat x(t), \hat \al(t),
\phi)\hat m_t(d\phi)dt
+c(\hat x(t^-),\hat \al(t^-))d\hat Z]\\ \aad\quad
=E_{x,\ell}^{\pi}\int_0^\tau\int_{\cal U}e^{-rt}[f( x(t), \al(t), \phi)
m_t(d\phi)dt
+c(x(t^-),\al(t^-))dZ].
\earray
$$
Thus, as $h \to 0$, $$J^h(x,\ell,\pi^h) \to J(x,\ell, \pi).$$ \qed

\begin{thm}\label{Value}
$V^h(x,\ell)$ and $V(x,\ell)$ are value functions
defined in {\rm \eqref{revalue}} and {\rm \eqref{value-numerical-purpose}},
respectively. Then $V^h(x,\ell)\to V(x,\ell)$ as $h\to0$.
\end{thm}
\para{Proof.}
First, to prove
\beq{cost1}
\limsup_hV^h(x,\ell)\leq V(x,\ell).
\eeq
Since $V(x,\ell)$ is
the maximizing cost function, for any
admissible
control $\pi\cd$,
$$J(x,\ell,\pi) \leq V(x,\ell).$$
Let $\wdt m^{h}\cd$ be an optimal relaxed control for
$\{\xi^{h}\cd\}$ and $\wdt \pi^h\cd=(\wdt m^{h}\cd, \wdt z^h\cd, \wdt g^h\cd)$.
That is,
$$ V^{h}(x,\ell) =J^{h}(x,\ell, \wdt \pi^{h})=\sup_{\pi^{h} }
J^h(x,\ell, \pi^{h}).$$ Choose a subsequence $\{\wdt h\}$ of
$\{h\}$ such that
$$\lim_{\wdt h\to 0 } V^{\wdt h} (x,\ell) = \limsup_{\wdt h\to 0}
V^{\wdt h} (x,\ell)=   \lim_{\wdt h\to 0 } J^{\wdt h}
(x,\ell,\wdt \pi^{\wdt h}).$$
Without loss of generality (passing to
an additional subsequence if needed), we may
assume that $(\xi^{\wdt
h}\cd, \al^{\wdt h}\cd, m^{\wdt h}\cd, w^{\wdt h}\cd, z^{\wdt h}\cd, g^{\wdt
h}\cd)$
converges weakly to $(x\cd, \al\cd, m\cd, w\cd, z\cd, g\cd  )$,
 where
$\pi\cd$ is an admissible related control. Then the weak convergence
and the Skorohod representation yield that \beq{low-bd} \limsup_{h}
V^h(x,\ell) = J(x,\ell,\pi) \le V(x,\ell).\eeq We proceed to
prove the reverse
 inequality.

 We claim that \beq{upp-bd} \liminf_{h} V^h(x,\ell) \ge V(x,\ell)
.\eeq Suppose that $\lbar m$ is an optimal control with Brownian
motion $w\cd$ such that $\lbar x\cd$ is the associated
trajectory. By the chattering lemma, given any $\gamma>0$, there are
an $\e>0$ and an ordinary control $u^\gamma\cd$ that takes only finite
many values, that $u^\gamma\cd$ is a constant on $[k \e, k\e +\e)$,
that $\lbar m^\gamma\cd$ is its relaxed control representation, that
$(\lbar x^\gamma\cd, \lbar m^\gamma\cd)$ converges weakly to $(x\cd,
\lbar m\cd)$, and that $J(x,\ell,\lbar \pi^\gamma) \ge V
(x,\ell) -\gamma.$

For each $\gamma>0$, and the corresponding $\e>0$ as in the chattering
lemma,  consider an optimal control problem as in (\ref{1d-state})
with piecewise constant on $[k \e, k\e +\e)$. For this controlled
diffusion process, we consider its $\gamma$-skeleton. By that we mean
we consider the process $(x^\gamma(k\e), m^\gamma(k\e))$. Let $\wdh
u^\gamma\cd$ be the optimal control, $\wdh m^\gamma\cd$ the relaxed
control representation, and $\wdh x^\gamma\cd$ the associated
trajectory. Since $\wdh m^\gamma\cd$ is optimal control,
$J(x,\ell,\wdh m^\gamma)\ge J(x,\ell,\lbar m^\gamma) \ge
V(x,\ell)-\gamma.$ We next approximate $\wdh u^\gamma\cd$ by a suitable
function of $(w\cd,\al\cd)$. Moreover, $ V^h (x,\ell)\ge
J^h(x,\ell,\lbar m^h)\to J(x,\ell,\lbar m^{\gamma,\theta})$
Thus,
$$\liminf_{h}
V^h (x,\ell)\ge J^h(x,\ell,\lbar m^h)\to
J(x,\ell,\lbar m^{\gamma,\theta}).$$ Using the result obtained in
\propref{approx}, $$\liminf_h V^h(x,\ell) \ge V(x,\ell)- 2\gamma.$$ The
arbitrariness of $\gamma$ then implies that $\liminf_{h} V^h(x,\ell)
\ge V(x,\ell).$

Using (\ref{low-bd}) and (\ref{upp-bd}) together with the weak
convergence and the Skorohod representation, we obtain the desired
result. The proof of the theorem is concluded. \qed

\section{Numerical Example}\label{sec:exm}

This section is devoted to a couple of examples. For simplicity,
we consider the case
that the discrete event has two states. That is,
the continuous-time Markov chain has two states. We approximate
the value functions in the case of the claim size distributions
are given. Proportional reinsurance and nonproportional
reinsurance are considered, respectively. These results are
compared to the numerical examples in \cite{AsmussenHT}.

\subsection{Proportional Reinsurance}
\begin{exm}\label{exp-pro} {\rm
The  generator of the Markov chain $\al(t)$ is
  $$
  Q=\left(
  \begin{array}{cc}
  -0.5 & 0.5\\
  0.5 & -0.5
  \end{array}\right),
  $$
and  $\M=\{1,2\}.$ The claim rate depends on the discrete state
with $\beta(1)=1$ and $\beta(2)=10$. Assume the claim size
distribution to be exponential with parameter 1. Then $E[Y]=1$ and
$E[Y^2]=2$. The \eqref{1d-state} follows \bea
\begin{cases}
dX(t)= \beta(\al(t)) u(t)dt + \sqrt{2\beta(\al(t))}u(t)dw(t)-dZ(t), \\
X(0^-)=x
\end{cases}
\eea where the retention level $u(t)$ is the regular control
parameter representing the fraction of the claim covered by the
cedent and $u(t) \in [0,1]$. Taking the discount rate $r=0.05$, we
compare the cost function in the case of the total expected
discounted value of all dividends until lifetime ruin mentioned in
\eqref{costfun1}
$$
J(x, \ell, \pi)= E_{x, \ell}\int_{[0,\tau]} e^{-rt} dZ(t).
$$
and the differential marginal yield to measure the instantaneous
returns accrued from irreversibly exerting the singular policy,
see \cite{Alvarez}.
$$
J(x, \ell, \pi)= E_{x, \ell}\int_{[0,\tau)} \hat \lambda e^{-rt-
\hat \lambda X(s)} dZ(t),
$$
where $\hat \lambda =1$. We obtained \figref{fig:nonpro-2} for this case.

\begin{figure}[htbp!]
\begin{center}
\mbox{\subfigure[\footnotesize Total expected discounted
value of all dividends versus initial surplus]
{\label{fig:nonpro-1-a}\epsfig{figure=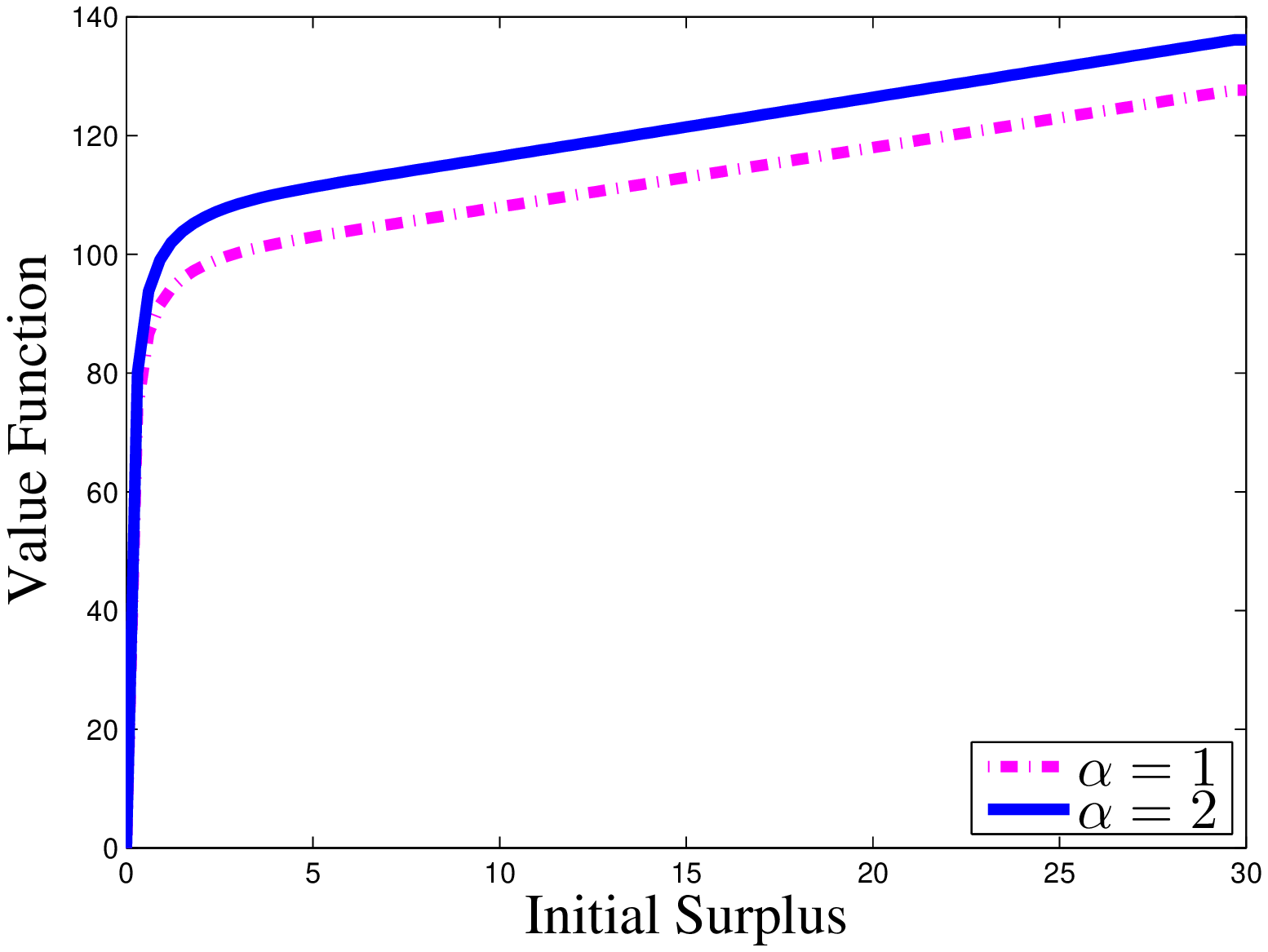,width=0.46\linewidth}}} \ \ \
\mbox{\subfigure[\footnotesize Differential marginal yield versus initial surplus]
{\label{fig:nonpro-1-b}\epsfig{figure=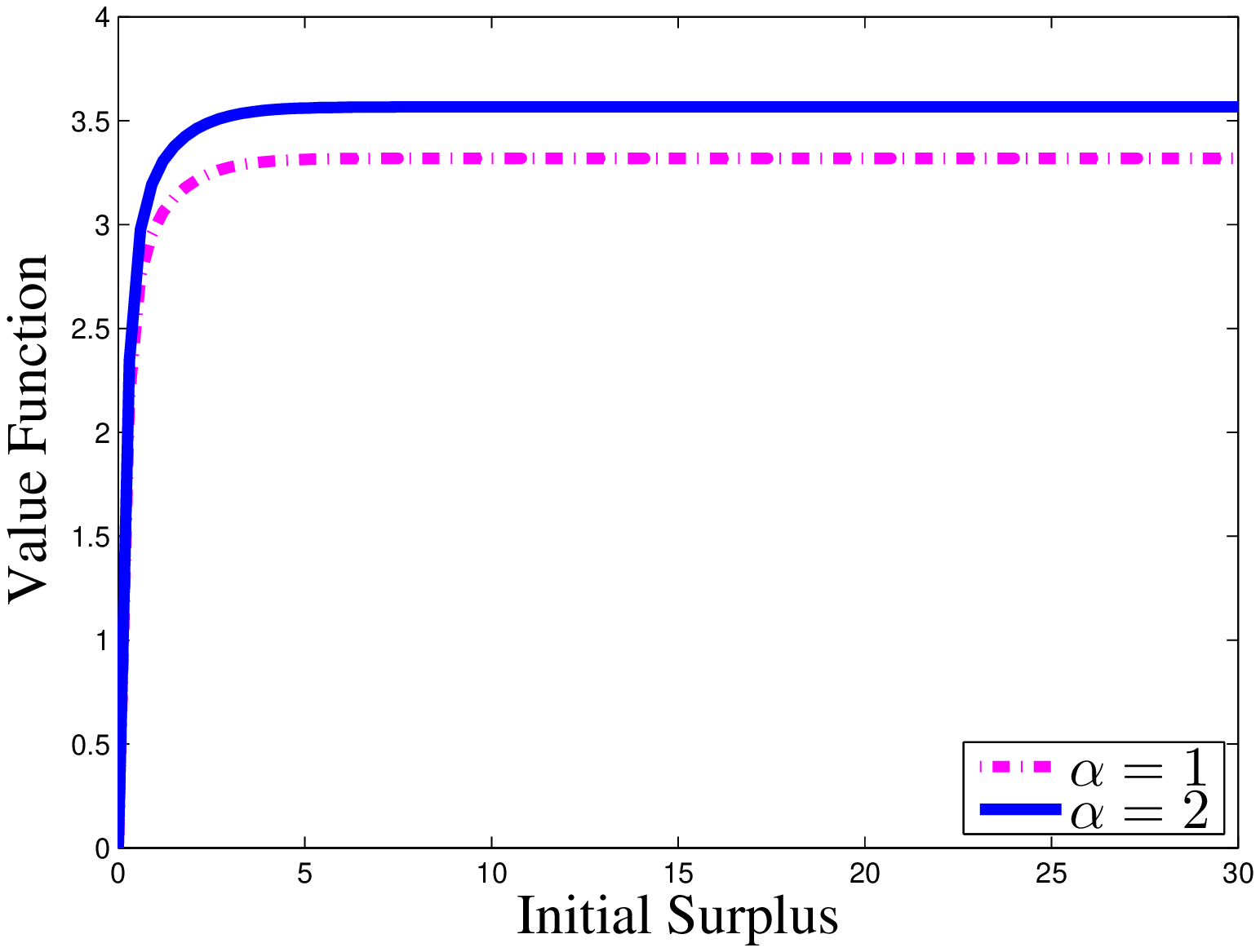,width=0.46\linewidth}}} \\
\mbox{\subfigure[\footnotesize Optimal reinsurance policy to total expected discounted
value of all dividends versus initial surplus]
{\label{fig:nonpro-1-c}\epsfig{figure=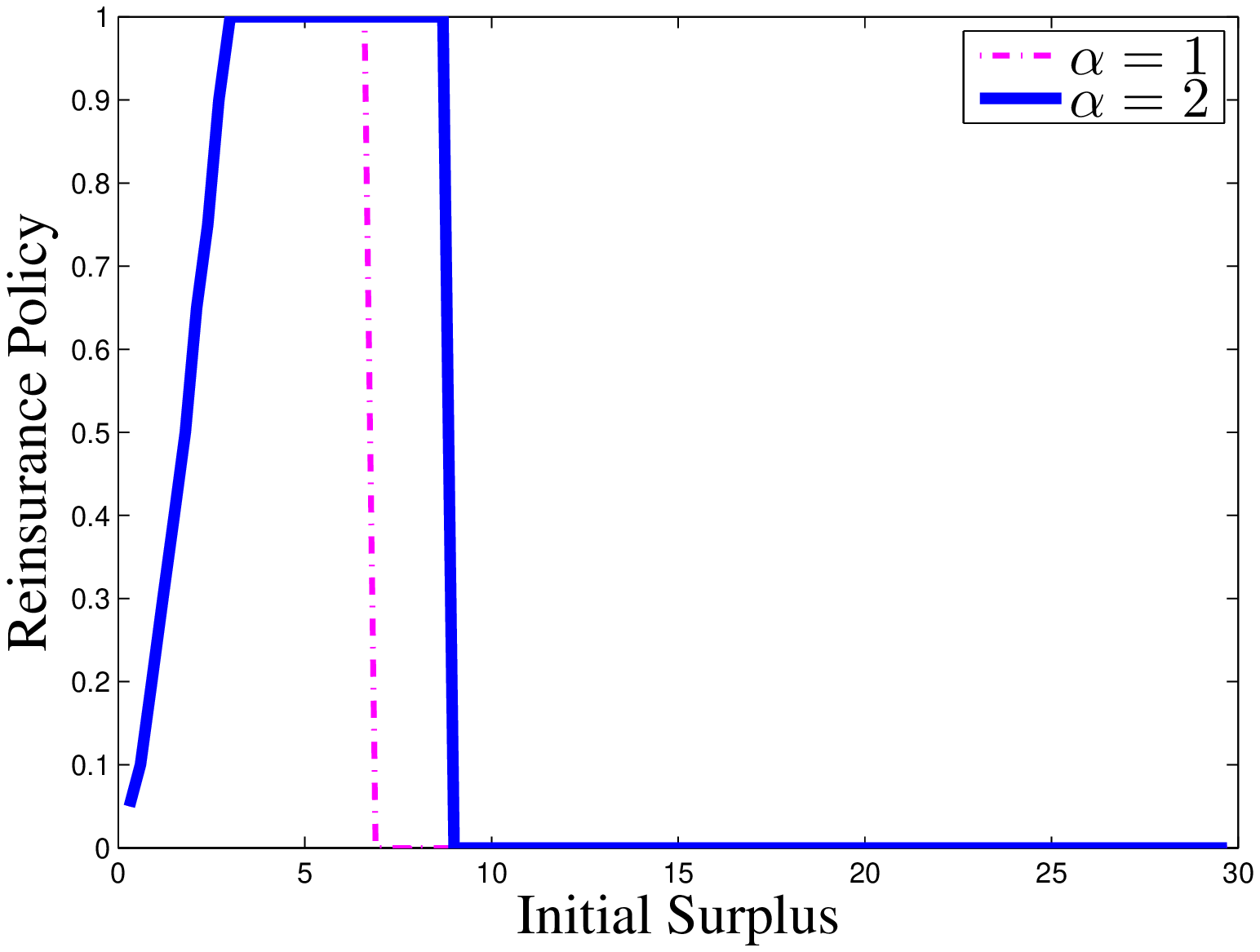,width=0.46\linewidth}}}\ \ \
\mbox{\subfigure[\footnotesize Optimal reinsurance policy to
differential marginal yield versus initial surplus]
{\label{fig:nonpro-1-d}\epsfig{figure=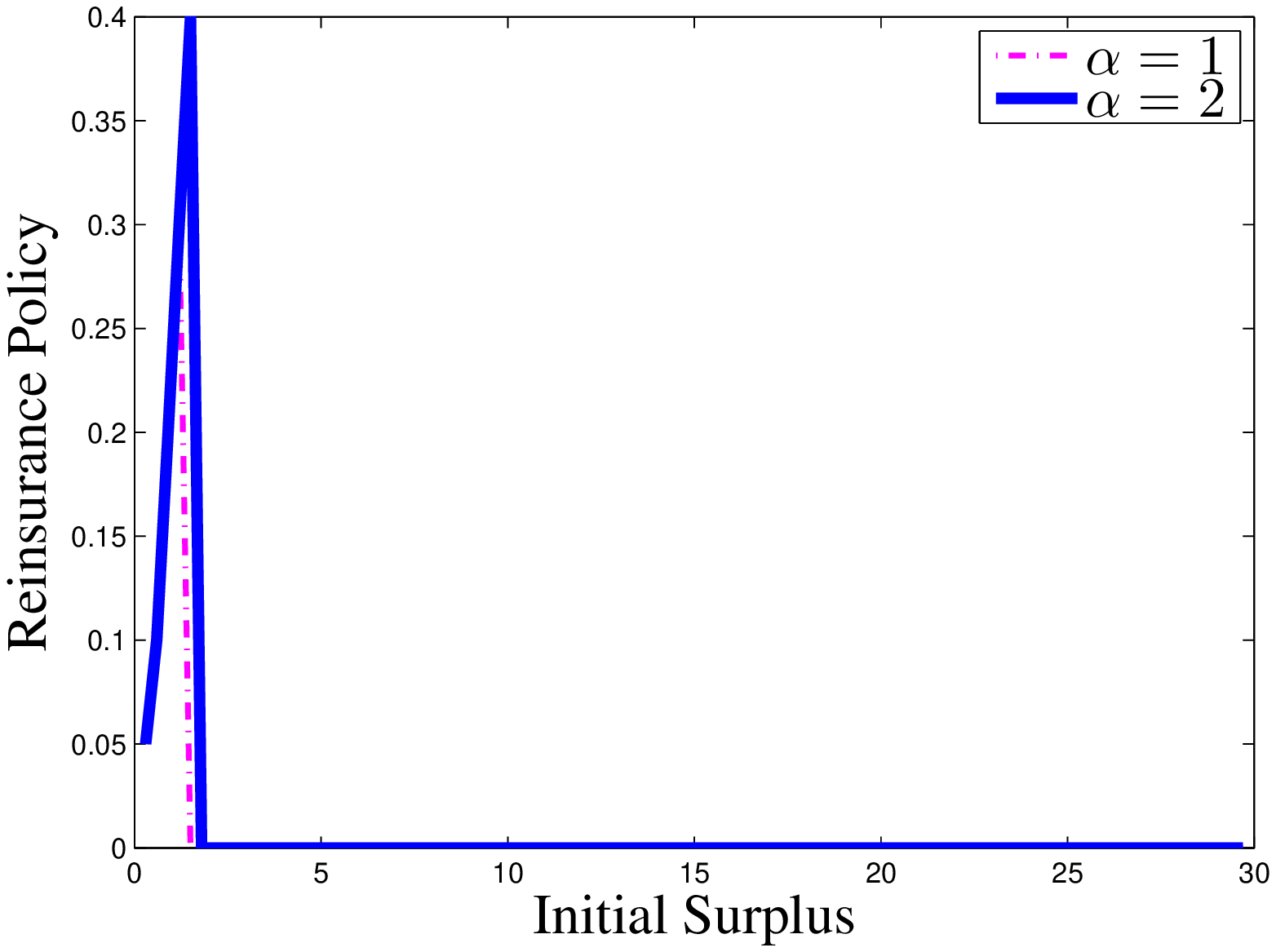,width=0.46\linewidth}}}
 \caption{Proportional reinsurance with exponential claim size distribution  with two regimes}
\end{center}\label{fig:nonpro-1}
\end{figure}

}
\end{exm}

\begin{exm}\label{uniform-pro} {\rm
In this example, the claim size distribution is assumed to be
uniform in $[0,1]$. Then $E[Y]=\frac{1}{2}$ and
$E[Y^2]=\frac{1}{3}$. Then the dynamic systems follows \bea
\begin{cases}
dX(t)=\frac{1}{2}\beta(\al(t))u(t)dt
+ \sqrt{\frac{1}{3}\beta(\al(t))}u(t)dw(t)-
dZ(t), \\
X(0^-)=x.
\end{cases}
\eea
Using the same data and 
payoff function
in \exmref{exp-pro}, we then obtained
\figref{fig:nonpro-2}.

\begin{figure}[htbp!]
\begin{center}
\mbox{\subfigure[\footnotesize Total expected discounted
value of all dividends versus initial surplus]
{\label{fig:nonpro-2-a}\epsfig{figure=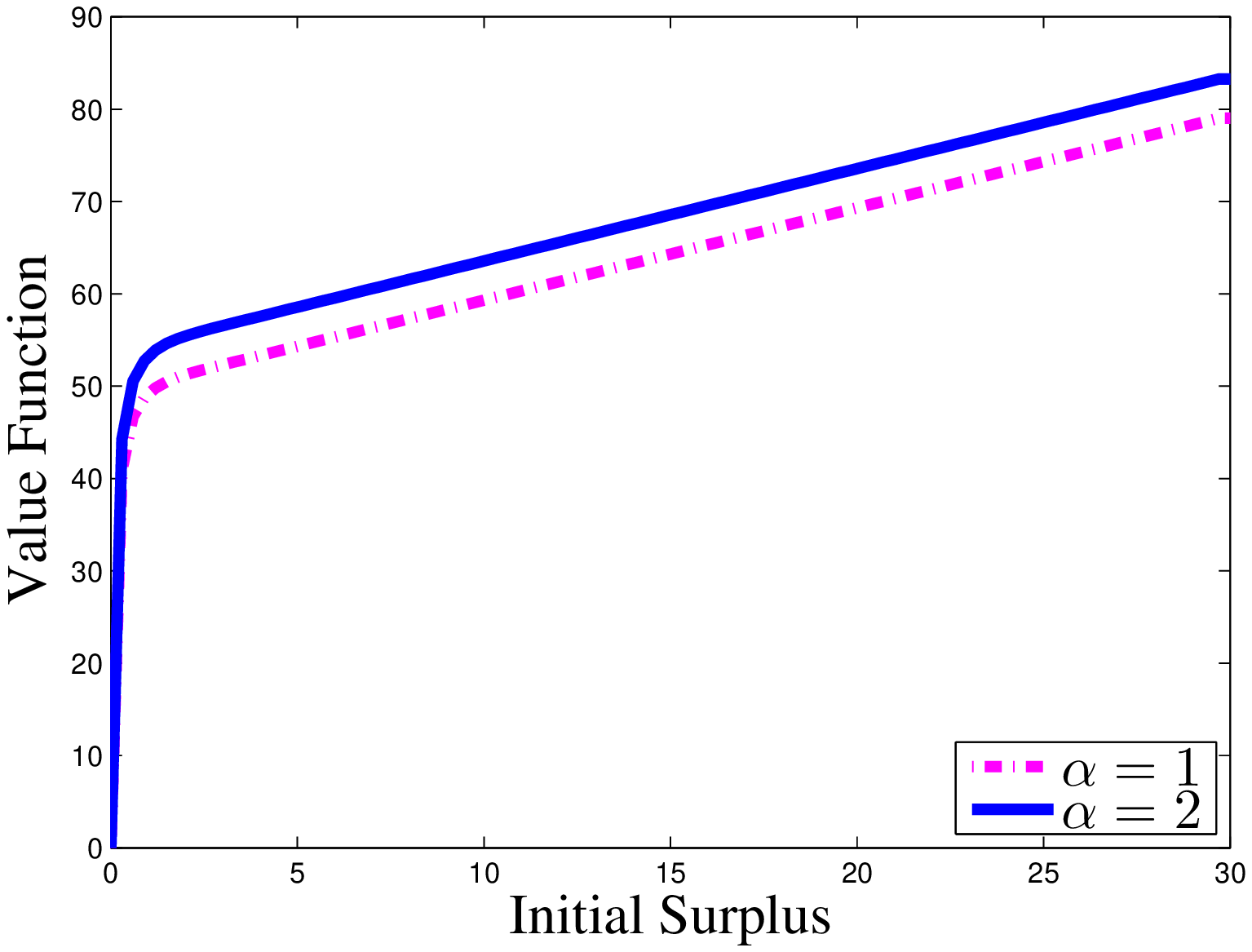,width=0.46\linewidth}}} \ \ \
\mbox{\subfigure[\footnotesize Differential marginal yield versus initial surplus]
{\label{fig:nonpro-2-b}\epsfig{figure=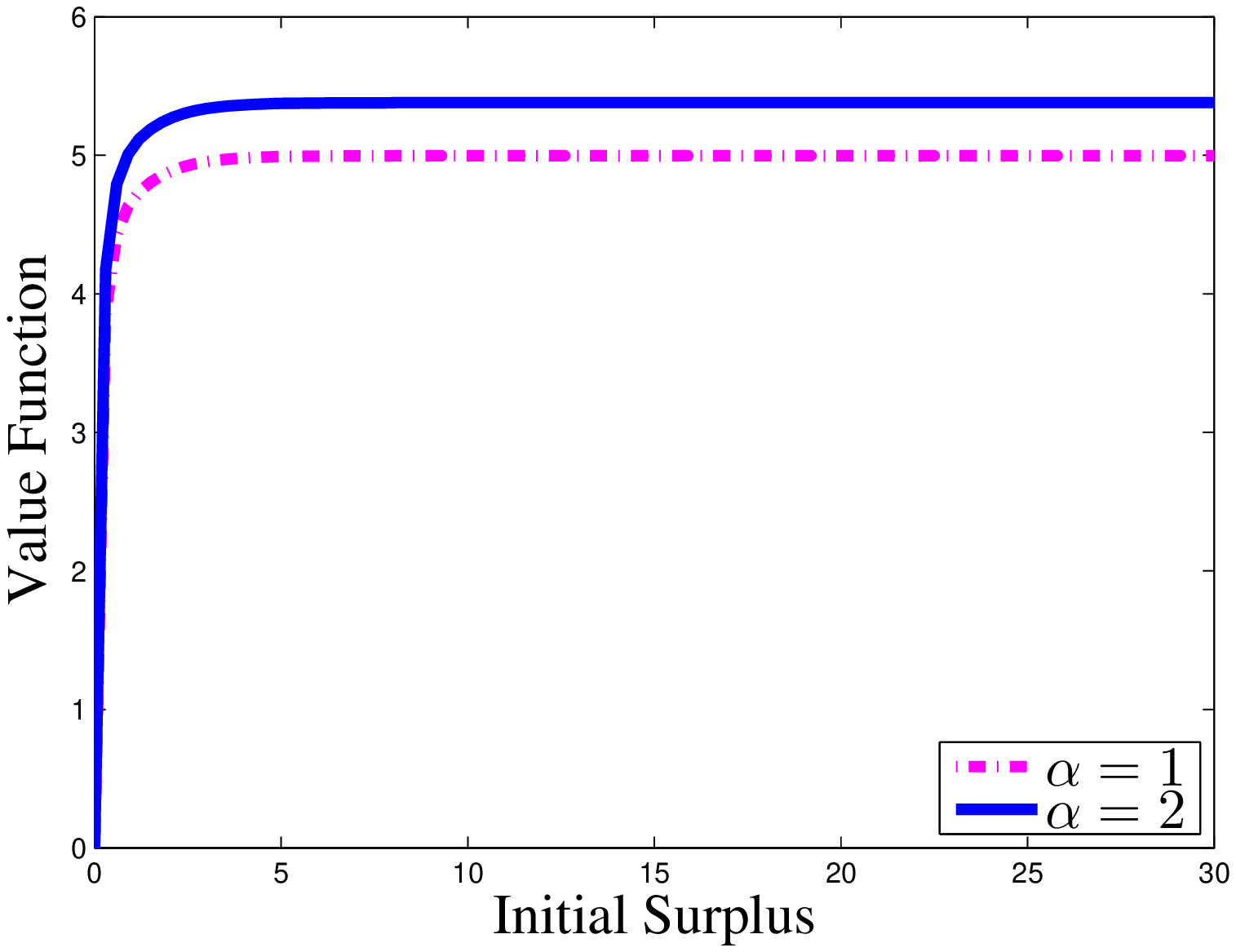,width=0.46\linewidth}}}\\
\mbox{\subfigure[\footnotesize Optimal reinsurance policy to total expected discounted
value of all dividends versus initial surplus]
{\label{fig:nonpro-2-c}\epsfig{figure=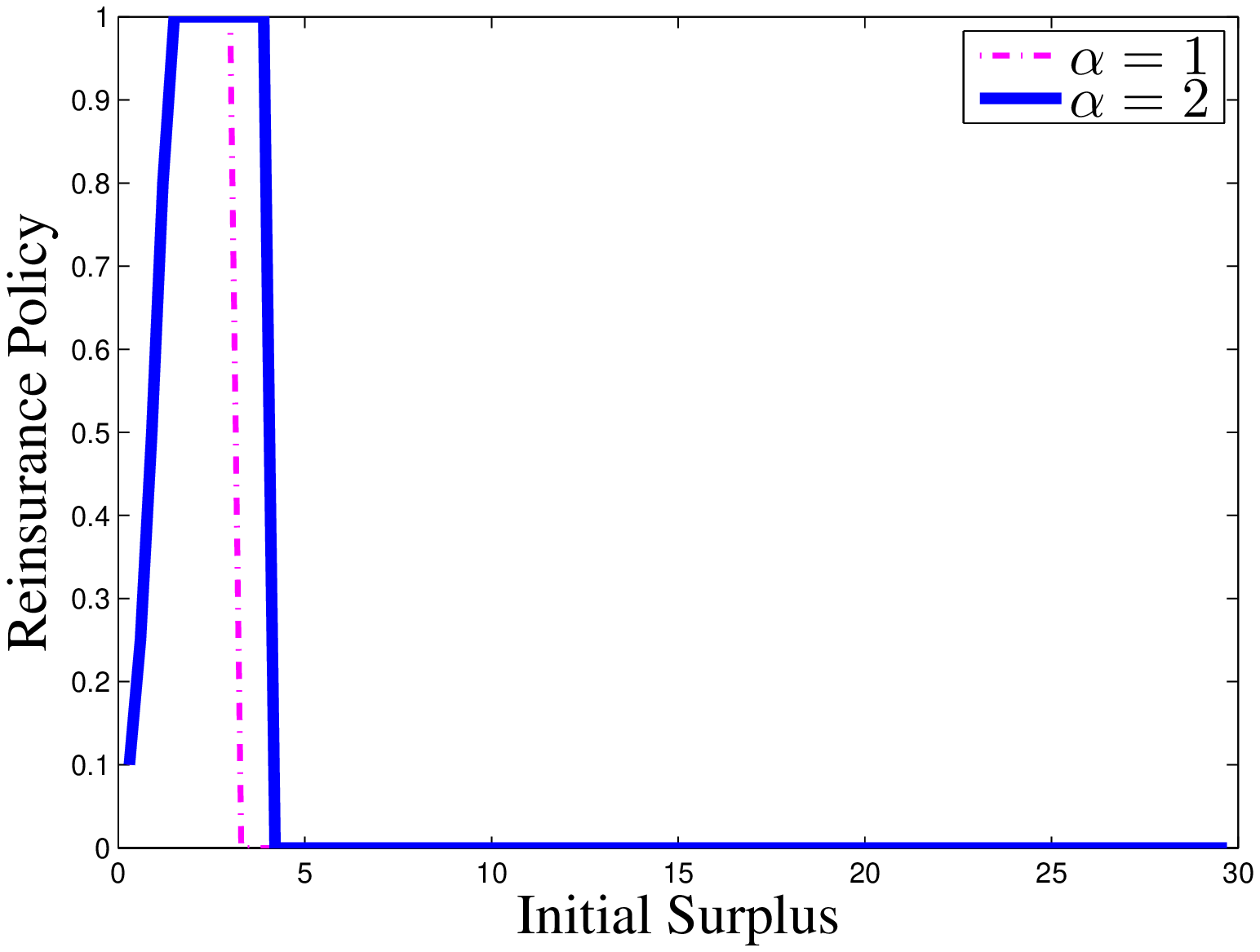,width=0.46\linewidth}}}\ \ \
\mbox{\subfigure[\footnotesize Optimal reinsurance policy to
differential marginal yield versus initial surplus]
{\label{fig:nonpro-2-d}\epsfig{figure=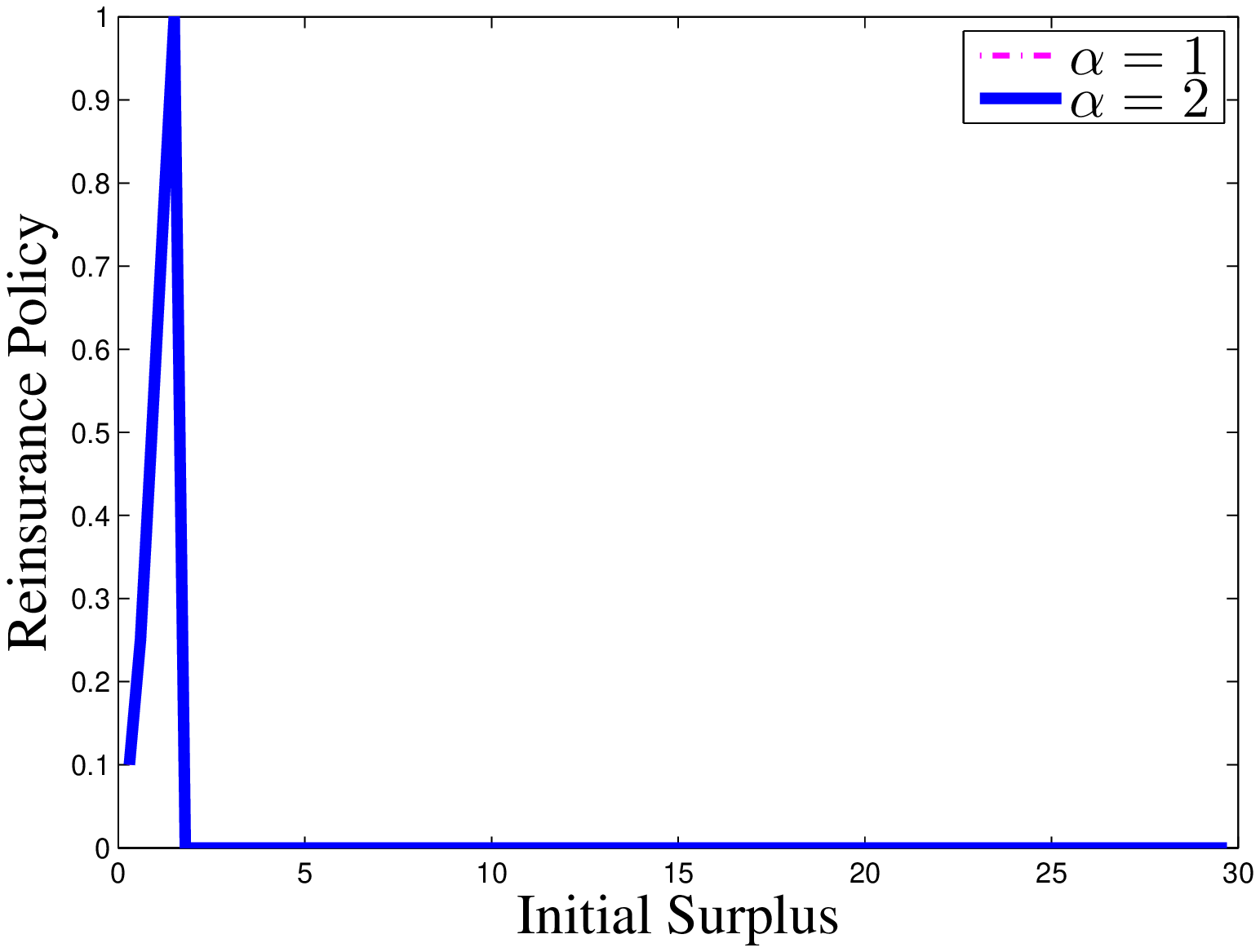,width=0.46\linewidth}}}
\caption{Proportional reinsurance with uniform claim size distribution  with two regimes}
\end{center}\label{fig:nonpro-2}
\end{figure}

}
\end{exm}
\subsection{Excess-of-Loss Reinsurance}

\begin{exm}\label{exp-non} {\rm
Comparing to \exmref{exp-pro} and \exmref{uniform-pro}, the
retention level $u(t)$ describes the maximal amount paid by the
cedent for each claim. Assume the claim size distribution to be
exponential with parameter 1 and $Q$, $\beta(1),\beta(2)$ and
payoff functions to be the same as those in Example 5.1.
Intuitively, the retention level cannot be arbitrarily large, then
we restrict the risk control set $U$ to be $[0,1]$. That is, the
retention level should not exceed the mean value of the
exponential distributed claim size. Following \eqref{moment}
$$ \barray \ad
E[Y^u]=\int_0^{u} e^{-x}dx =1 - e^{-u},\\ \ad
E[(Y^{u})^2]= \int_0^{u}
2x e^{-x}dx =2[1 - e^{-u}(1+u)]
.\earray
$$
Then the dynamic systems satisfy
$$
\left \{ \barray \ad
dX(t)=\beta(\al(t))[1 - e^{-u(t)}]dt+ \sqrt{2\beta(\al(t))[1 - e^{-
u(t)}(1+u(t))]} dw(t)-dZ(t), \\ \ad
X(0^-)=x;
\earray \right.
$$
see \figref{fig:pro-1} for this case.

\begin{figure}[htbp!]
\begin{center}
\mbox{\subfigure[\footnotesize Total expected discounted
value of all dividends versus initial surplus]
{\label{fig:pro-1-a}\epsfig{figure=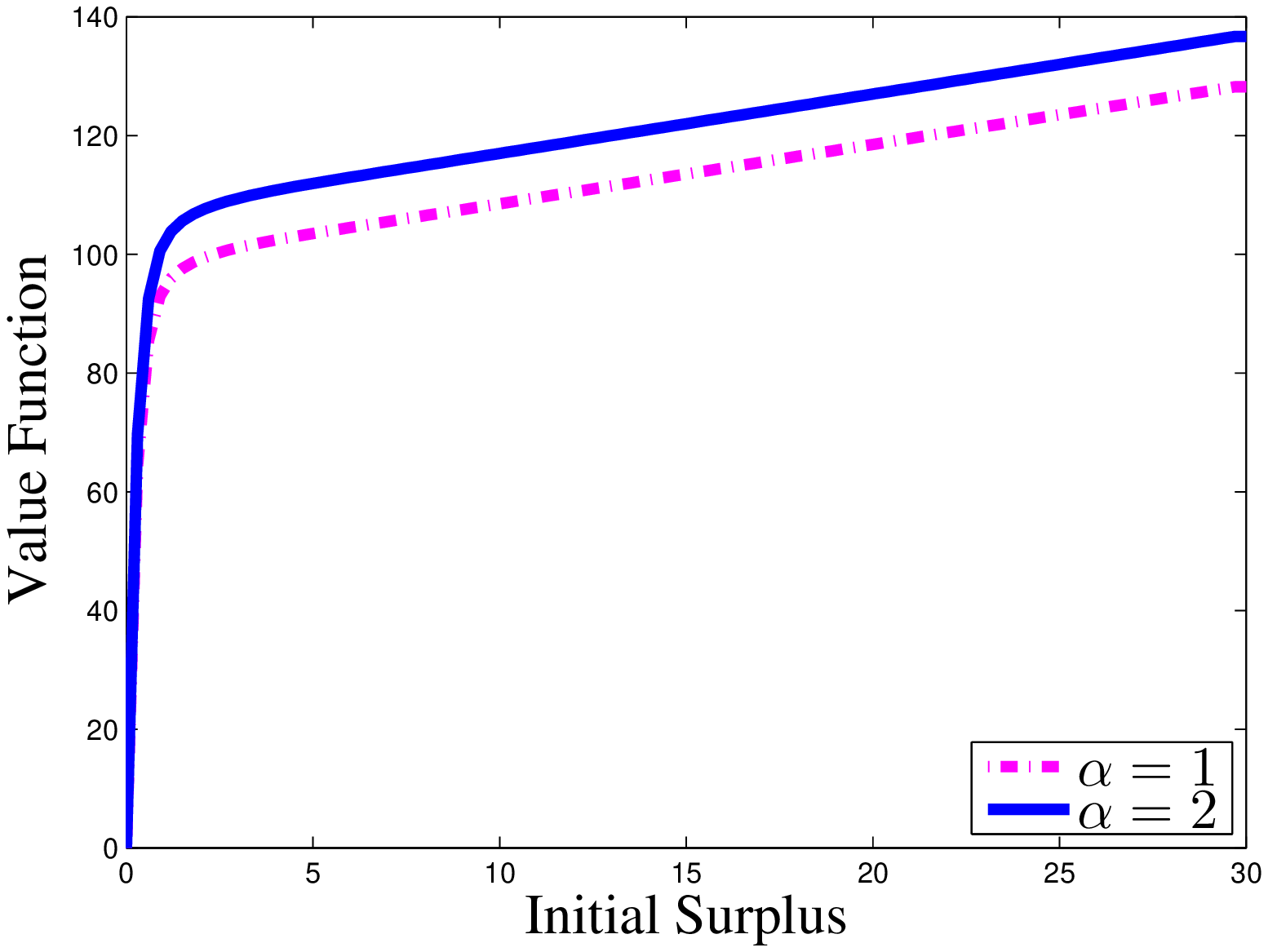,width=0.46\linewidth}}}\ \ \
\mbox{\subfigure[\footnotesize Differential marginal yield versus initial surplus]
{\label{fig:pro-1-b}\epsfig{figure=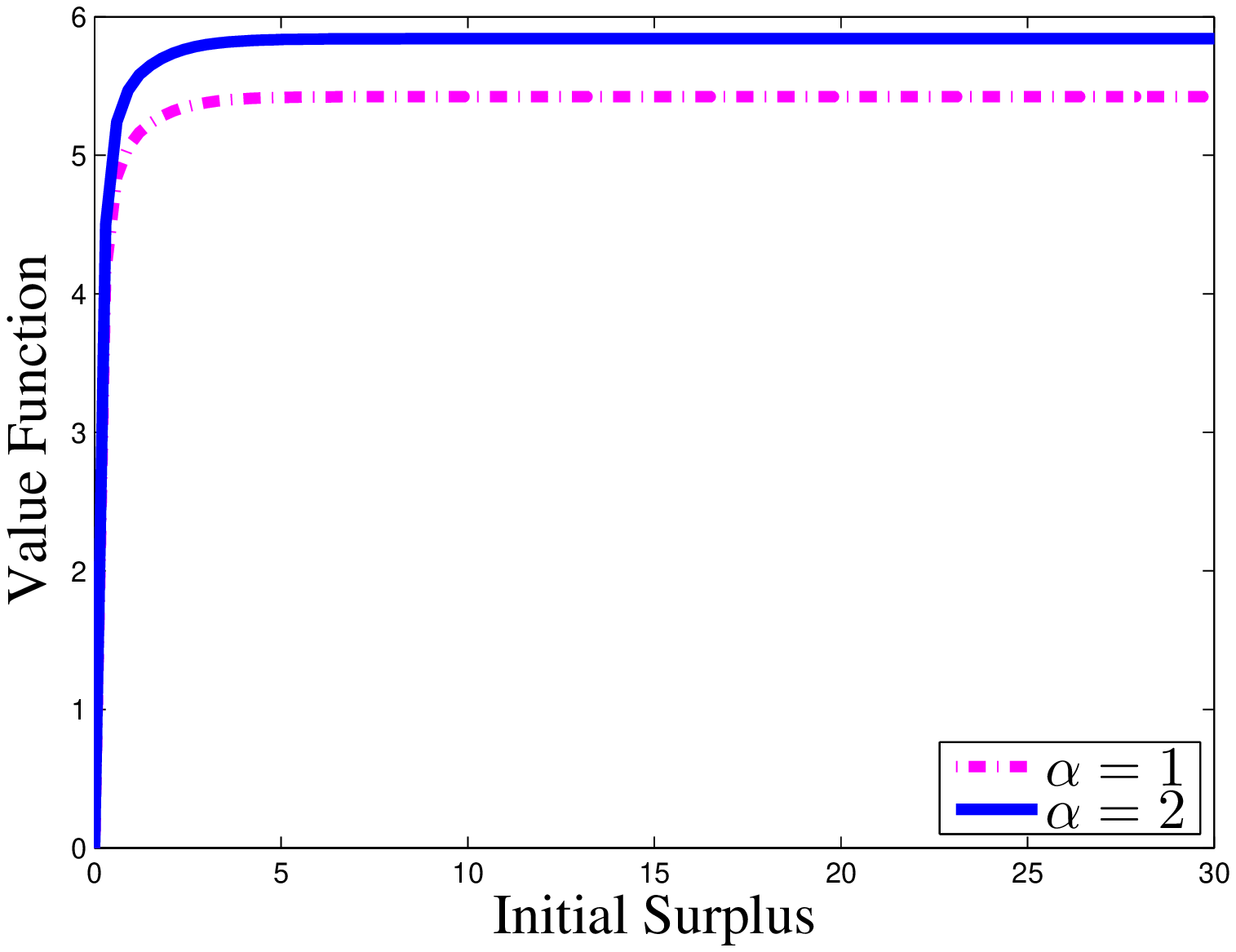,width=0.46\linewidth}}}
\end{center}\label{fig:pro-1}
\mbox{\subfigure[\footnotesize Optimal reinsurance policy to total expected discounted
value of all dividends versus initial surplus]
{\label{fig:pro-1-c}\epsfig{figure=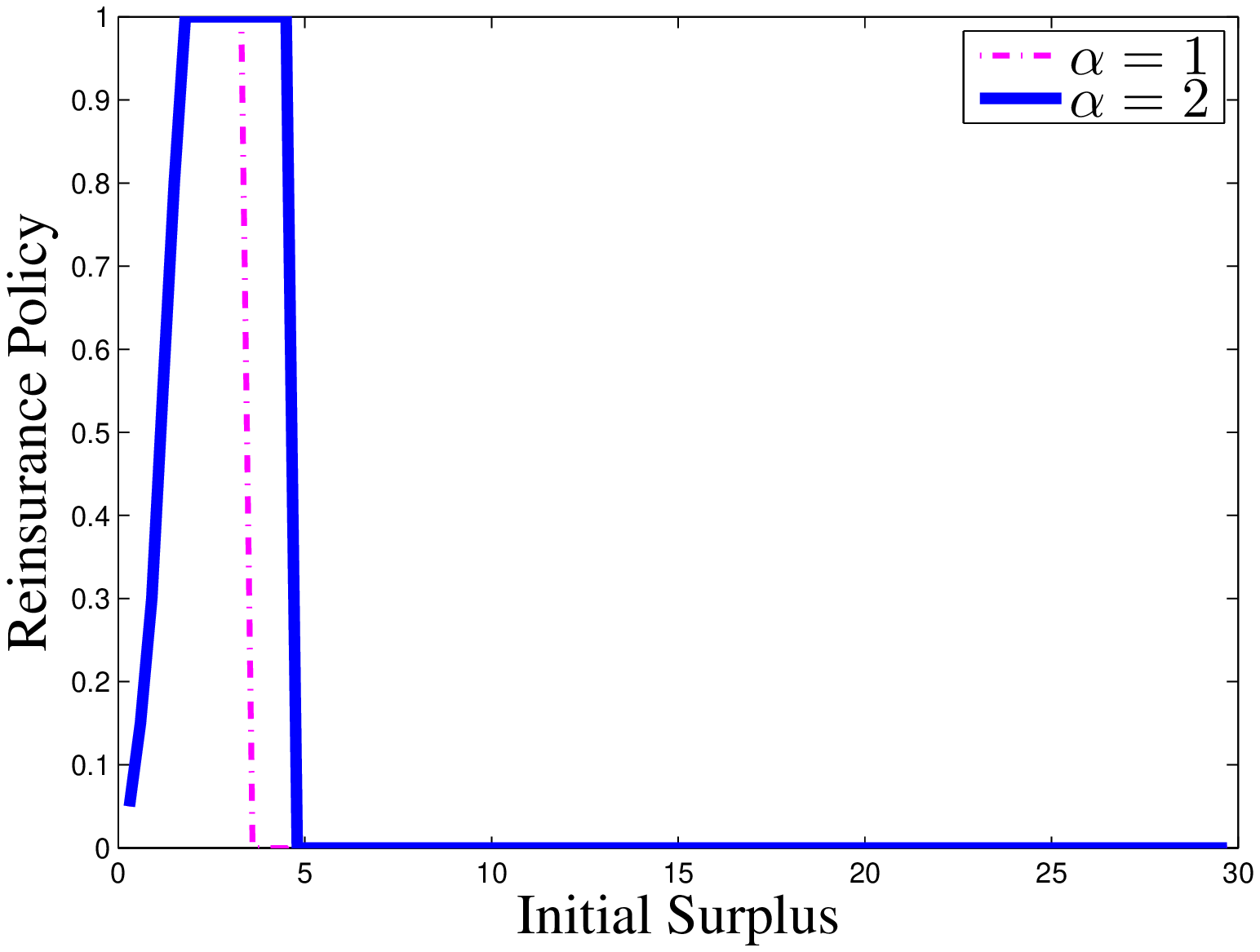,width=0.46\linewidth}}}\ \ \
\mbox{\subfigure[\footnotesize Optimal reinsurance policy to
differential marginal yield versus initial surplus]
{\label{fig:pro-1-d}\epsfig{figure=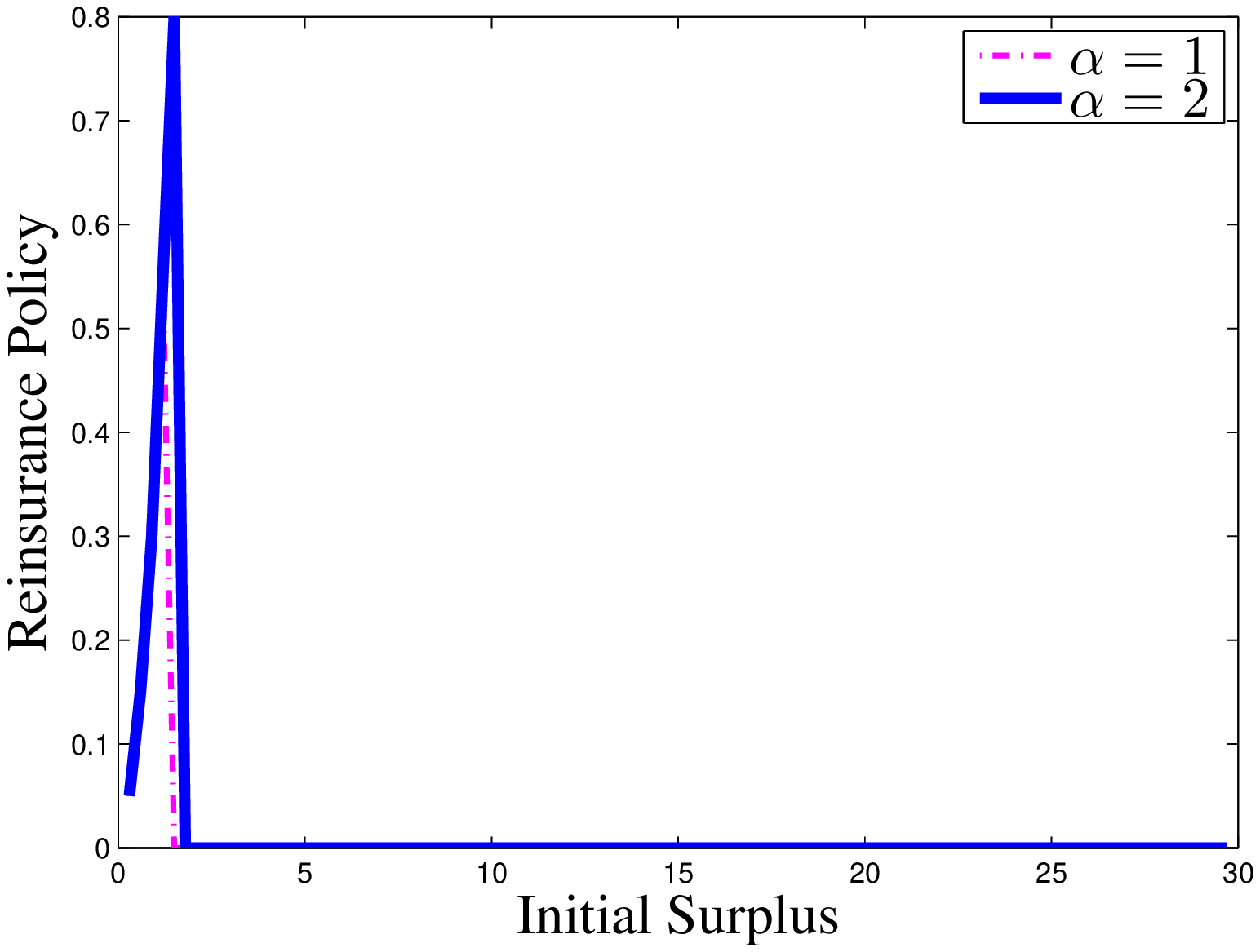,width=0.46\linewidth}}}
\caption{Excess-of-loss reinsurance with exponential claim size distribution with
two regimes}
\end{figure}

}
\end{exm}

\begin{exm}\label{uniform-non} {\rm
Assume the claim size distribution to be uniform in $[0,1]$.
Similarly, we obtain
$$ \barray \ad
E[Y^u]=\int_0^{u} (1-x)dx =u-\frac{u^2}{2},\\ \ad
E[(Y^{u})^2]= \int_0^{u}
2x(1-x)dx =u^2\Big[1-\frac{2u}{3}\Big].
\earray
$$
Hence, the dynamic systems satisfy
$$
\left \{ \barray \ad
dX(t)=\Big[\beta(\al(t))(u(t)-\frac{u(t)^2}{2})\Big]dt+
\sqrt{\beta(\al(t))u(t)^2\Big[1-\frac{2u(t)}{3}\Big]} dw(t)-dZ(t), \\ \ad
X(0^-)=x.
\earray \right.
$$
Let the risk control set $U=[0,1]$. We obtain \figref{fig:pro-2}
in this case.

\begin{figure}[htbp!]
\begin{center}
\mbox{\subfigure[\footnotesize Total expected discounted
value of all dividends versus initial surplus]
{\label{fig:pro-2-a}\epsfig{figure=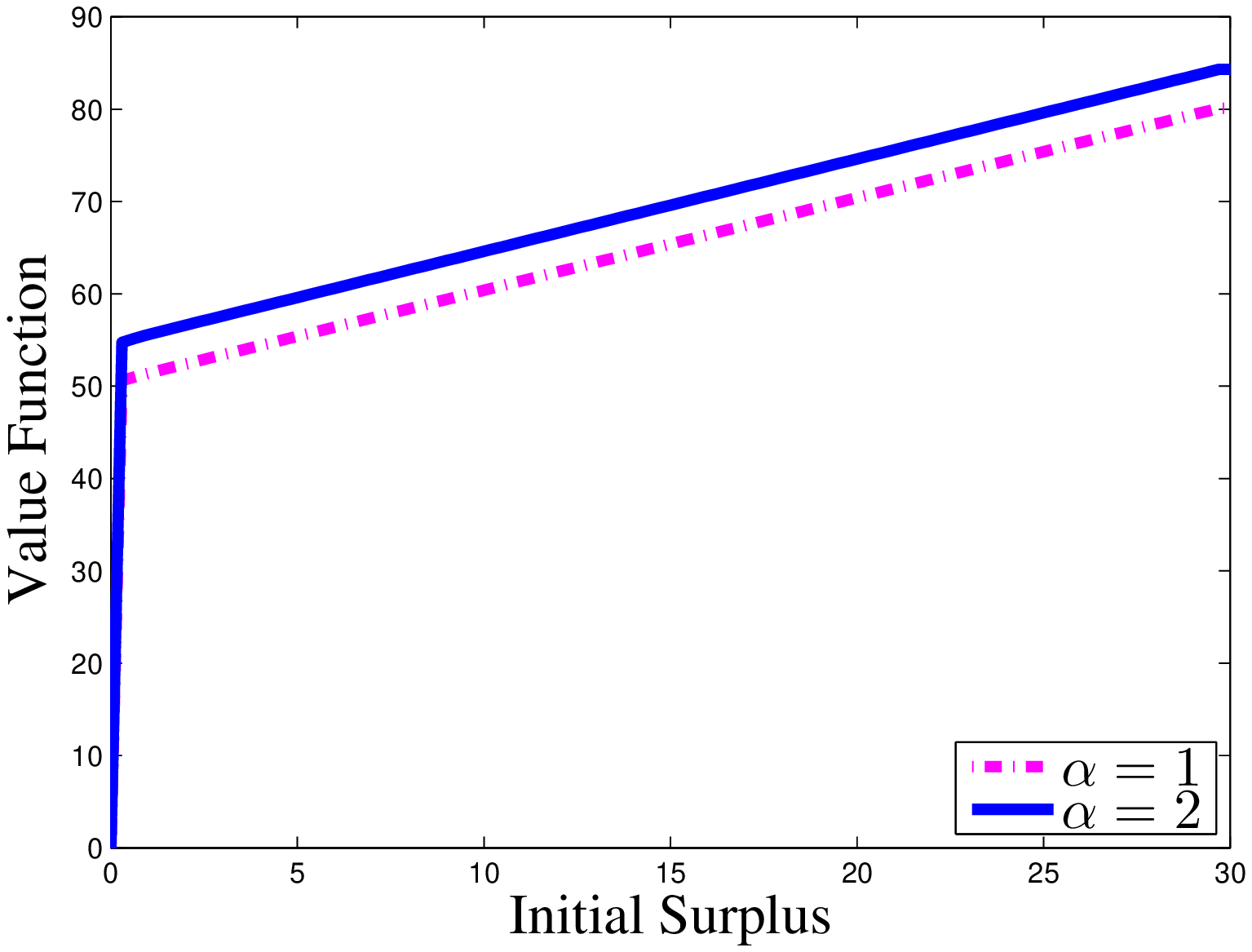,width=0.46\linewidth}}}\ \ \
\mbox{\subfigure[\footnotesize Differential marginal yield versus initial surplus]
{\label{fig:pro-2-b}\epsfig{figure=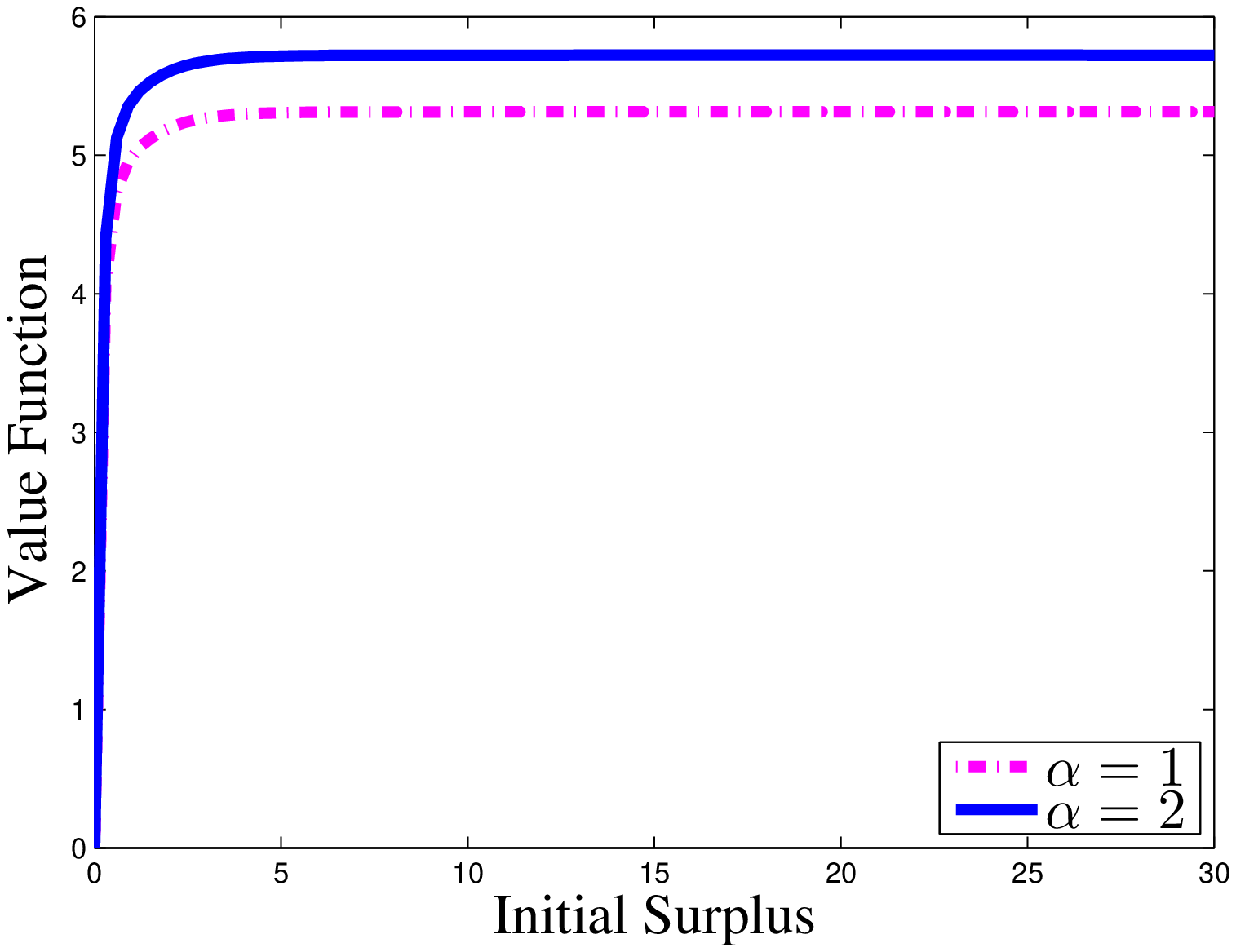,width=0.46\linewidth}}}
\mbox{ \subfigure[\footnotesize Optimal reinsurance policy to total expected discounted
value of all dividends versus initial surplus]
{\label{fig:pro-2-c}\epsfig{figure=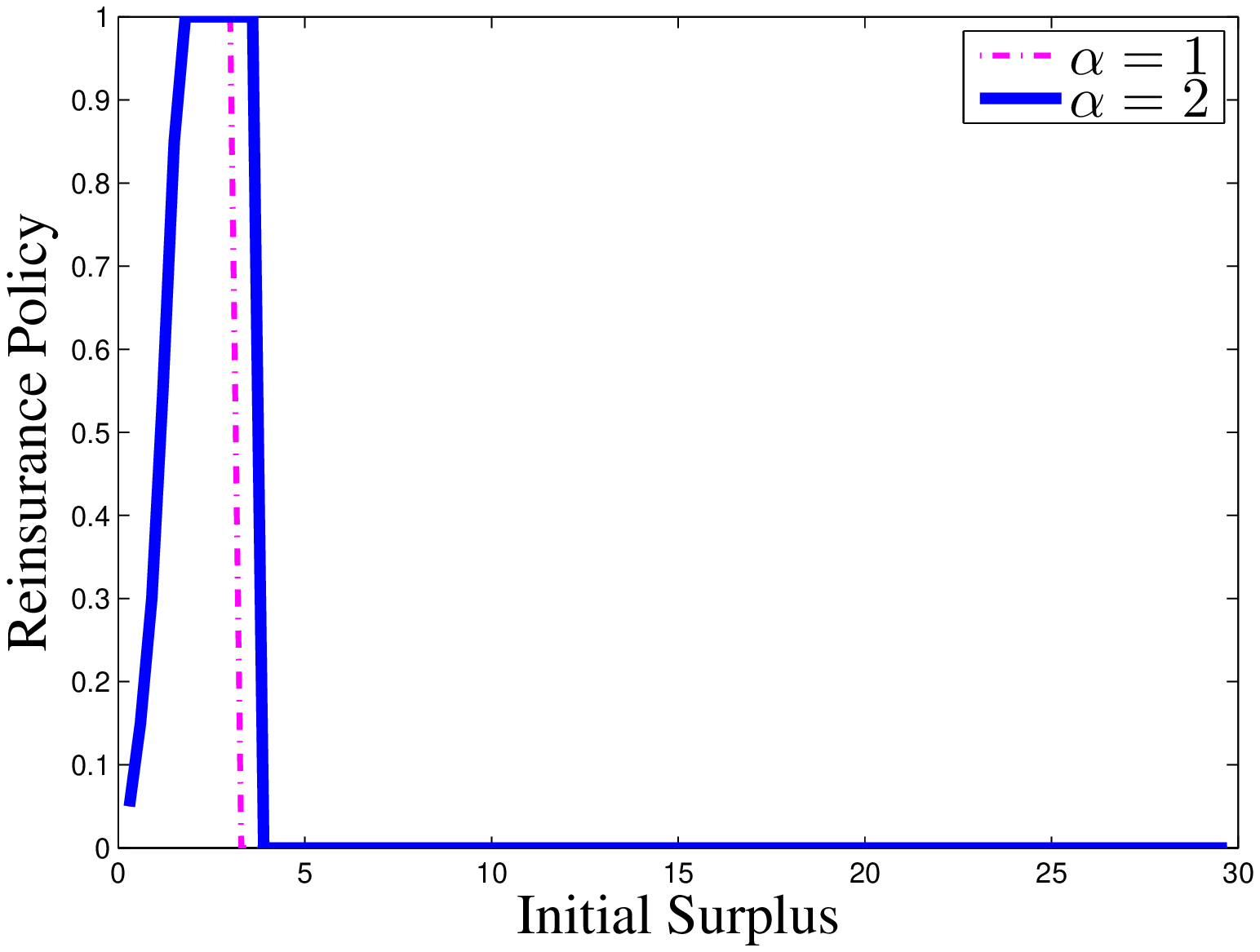,width=0.46\linewidth}}}\ \ \
\mbox{\subfigure[\footnotesize Optimal reinsurance policy to
differential marginal yield versus initial surplus]
{\label{fig:pro-2-d}\epsfig{figure=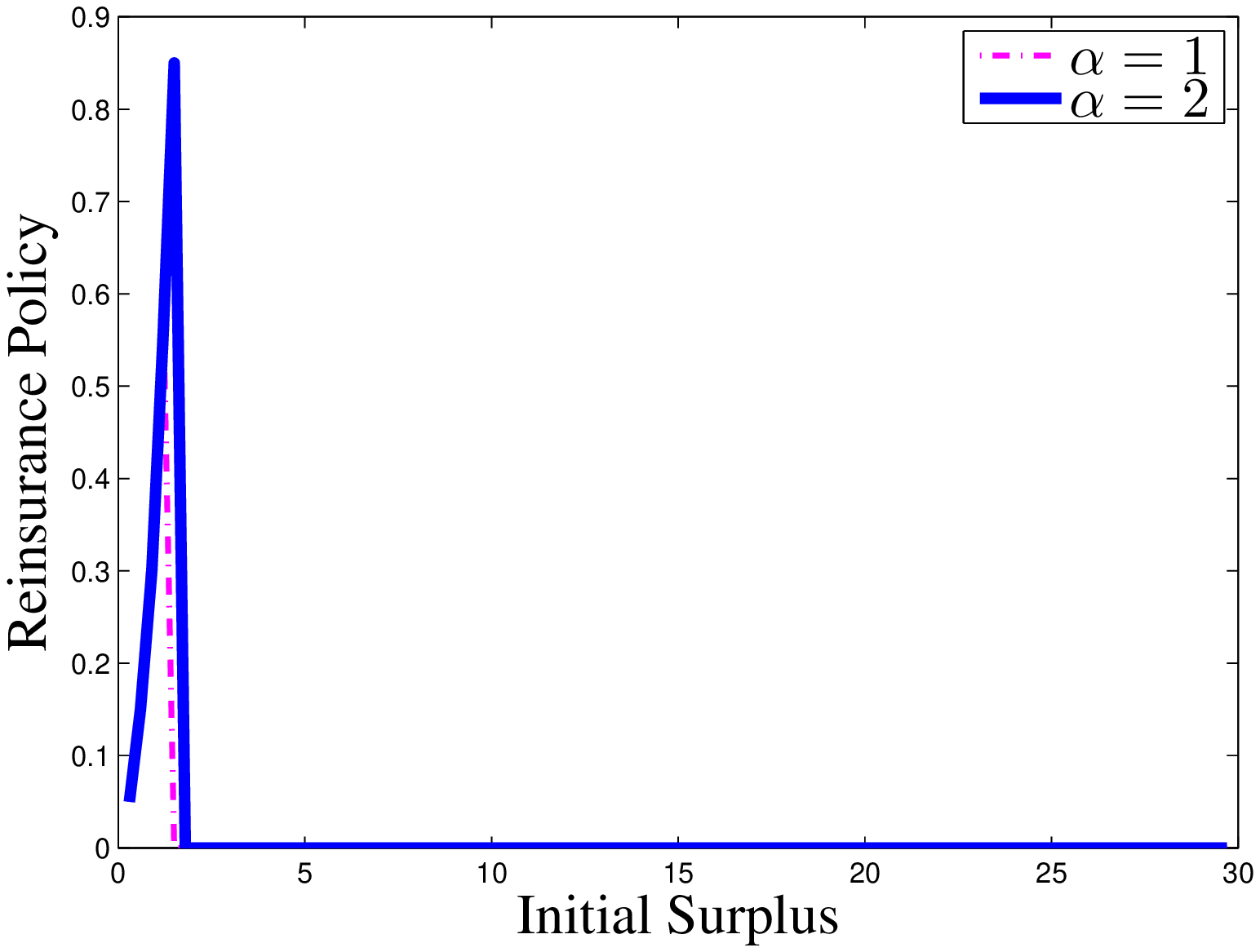,width=0.46\linewidth}}}\ \ \
\caption{Excess-of-loss reinsurance with uniform claim size distribution  with two regimes}
\end{center}\label{fig:pro-2}
\end{figure}

}
\end{exm}

All of the figures contain two lines since we consider the
two-regime case. Figures \subref{fig:nonpro-1-a}, 
\subref{fig:nonpro-2-a}, 
\subref{fig:pro-1-a} and 
\subref{fig:pro-2-a} show that the value function is concave and
the dividend payout strategy is a barrier strategy. It is clear
that if the surplus is higher than some barrier level, the extra
surplus will be paid as the dividend, with the same time the value
functions increase with 
 unity slope.

Regarding 
the reinsurance
policy, 
it is 
demonstrated in Figures \subref{fig:nonpro-1-c}, \subref{fig:nonpro-2-c}, \subref{fig:pro-1-c}, and \subref{fig:pro-2-c}
that both the proportional reinsurance and excess-of-loss
reinsurance increase at first, maintain the highest rate in an interval, and
decrease sharply to zero at a threshold to maximize the total expected discounted
value of all dividends. From Figures \subref{fig:nonpro-1-d}, \subref{fig:nonpro-2-d}, \subref{fig:pro-1-d}, \subref{fig:pro-2-d},
in the case of maximizing the differential marginal yield,
it is shown that both the proportional reinsurance and excess-of-loss
reinsurance have similar trend comparing to the case of maximizing
the total expected discounted value of all dividends, except that
there are not the interval to hold the highest reinsurance rate.
Furthermore, it is shown that in both regimes,
there exists a free boundary (barrier) that separates two regions
where
the regular control or singular control is dominant.
Also, the barrier levels are  different in different regimes due
to the Markov switching.

In addition, we compare the values for proportional reinsurance and excess-of-loss
reinsurance
with exponential claim size distribution in Table \ref{table:exp}.
At the level of initial surplus $x=30$, we 
 compare 
the corresponding values 
in two regimes. Similarly we have the
comparison in Table \ref{table:uniform}
for the uniform claim size distribution.

\begin{table}[h!t|b]
\begin{center}
\begin{tabular}{|c|c|c|}
  \hline
  Reinsurance type           & $\al=1$       & $\al=2$       \\ \hline
  proportional reinsurance   & 127.661229    & 136.139963    \\ \hline
  excess-of-loss reinsurance & 128.207117    & 136.686110    \\ \hline
 \end{tabular}
\caption{$V(30, \al)$ with exponential claim size distribution for
proportional reinsurance and excess-of-loss reinsurance}\label{table:exp}
\end{center}
\end{table}

\begin{table}[h!t|b]
\begin{center}
\begin{tabular}{|c|c|c|}
  \hline
  Reinsurance type           & $\al=1$       & $\al=2$       \\ \hline
  proportional reinsurance   & 79.010314     & 83.256482    \\ \hline
  excess-of-loss reinsurance & 80.097716     & 84.302264    \\ \hline
 \end{tabular}
\caption{$V(30, \al)$ with uniform claim size distribution for
proportional reinsurance and excess-of-loss reinsurance}\label{table:uniform}
\end{center}
\end{table}

From Tables \ref{table:exp} and   \ref{table:uniform}, we see that $V(30,
\al)$ of excess-of-loss reinsurance
are bigger than that of proportional reinsurance in both of the two regimes.
That is,
we can conclude that the excess-of-loss reinsurance is
more profitable than proportional reinsurance under the same condition. This is
consistent with the one regime case in \cite{AsmussenHT}. Finally, the numerical
method can treat complicate cost functions such as the marginal yield, which
is another advantage of the numerical solutions.

\section{Further Remark}\label{sec:rmk}

In this work, we have developed a numerical approximation scheme
to maximize the payoff function of the total discounted dividend
paid out until the lifetime of ruin. A generalized formulation of
reinsurance and dividend pay-out strategy is presented. Although
one could derive the associated system of QVIs by using the usual
dynamic programming approach together with the use of properties
of regime-switchings, solving for the mixed regular-singular
control problem analytically is very difficult. As an alternative,
we presented a Markov chain approximation method using mainly
probabilistic methods. For the singular control part, a technique
of time rescaling is used. In the actual computation, the optimal
value function can be obtained by using the value or policy
iteration methods. Examples of proportional and excess-of-loss
reinsurance are presented with more complicated payoff functions.

\appendix
\section{Appendix:
Proof of \propref{approx}}
The proof is similar in spirit to \cite{YinJJ}
however regime-switching is included. The technique are originated
from the work of \cite{Kushner-D}.
For simplicity,
We divide the proofs into several steps.
First, for any $\varsigma>0$, by \thmref{chattering}, there are $\e>0$,
a finite set $U^\varsigma \subset U$, and a probability space on which
are defined a solution in the stochastic differential equation in
(\ref{1d-state}). Thus, we have $(x^\varsigma\cd, \alpha^\varsigma\cd,
u^\varsigma\cd, w^\varsigma\cd)$, where $U^\varsigma\cd$ is $U^\varsigma$-valued
and constant on $[n \e, n\e+\e)$. Moreover, $(
x^\varsigma\cd,\alpha^\varsigma\cd, m^\varsigma\cd, w^\varsigma\cd)$ converges
weakly to $(X\cd,\alpha\cd,m\cd,w\cd)$, the solution of the
differential equation in (\ref{1d-state}). This further implies that
$ \limsup_\varsigma | J^{m^\varsigma}_{x,\iota}
\cd- J^m_{x,\iota}\cd| \le
\e_{\wdt \varsigma}$ with $\e_{\wdt \varsigma}$ satisfying $\e_{\wdt
\varsigma}\to
0$ as $\wdt \varsigma\to 0$.

Next, consider a $ u^\varsigma\cd$  for $\varsigma$
sufficiently small. Let $0< \theta <\e$.
For $\phi\in U^\varsigma$, define
the function $F_{n,\theta}$ as the regular conditional probability \bea
\ad F_{n,\theta} (\phi, u^\varsigma(i \e), i< n, w^\varsigma( p\theta), p \theta
\le n\e) = P(u^\varsigma( n\e)=\phi| u^\varsigma(i \e), i< n,
w^\varsigma(p\theta) , p \theta \le n\e).\eea The uniqueness of the
solution of the wealth equation or the associated martingale
problem
implies that the law of
$(x^\varsigma,\alpha^\varsigma\cd,m^\varsigma\cd,w^\varsigma\cd)$
is determined
by
the law of $(\alpha^\varsigma\cd, m^\varsigma\cd,
w^\varsigma\cd)$ since the
$\sigma$-algebra determined by $\{u^\varsigma(i\e), i <n,
w^\varsigma(p\theta), \alpha^\varsigma(p\theta), p\theta \le n\e\}$ increases to
the $\sigma$-algebra determined by $\{ u^\varsigma(i \e), i < n,
w^\varsigma(\tau), \alpha^\varsigma(\tau), \tau \le n\e\}$ as $\theta\to 0$,
we can show that for each $n$, $\phi$, and $\e$,
$F_{n,\theta} (\phi, u^\varsigma(i \e), i< n,
w^\varsigma( p\theta),\alpha^\varsigma(p\theta), p \theta \le n\e)  \to
P(u^\varsigma(
n\e)=\phi| u^\varsigma(i \e), i< n, w^\varsigma( \tau)
,\alpha^\varsigma(\tau), \tau \le n\e)$
with probability one as
$\theta\to 0$.

For $w^{\varsigma ,\theta}\cd$, define the control $u^{\varsigma, \theta}\cd$
by the conditional probability given in $F_{n,\theta}$
with $\varsigma$ replaced by $\varsigma ,\theta$. Owing to the
construction of the control law, as $\theta\to 0$, $(\alpha^{\varsigma,\theta}
\cd,m^{\varsigma,\theta}\cd, w^{\varsigma, \theta}\cd)$ converges weakly to $
(\alpha^{\varsigma}\cd, m^\varsigma\cd, w^\varsigma\cd)$.
We can further
show  $(x^{\varsigma,\theta}\cd, \alpha^{\varsigma,\theta}\cd,
m^{\varsigma,\theta}\cd,
w^{\varsigma ,\theta}\cd)$ converges weakly to $(x^\varsigma\cd,
\alpha^\varsigma\cd, m^\varsigma\cd, w^\varsigma\cd)$ as $\theta\to 0$, and
moreover $x^{\varsigma,\theta}\cd$ converges weakly to $x^\varsigma\cd$. Thus
$| J^{m^{\varsigma, \theta}}_{x,\iota}\cd-
J^{m^\varsigma}_{x,\iota}\cd | \le g_1(\theta)$,
where $ g_1(\theta)
\to 0$ as $\theta\to 0$.

For $\Delta>0$, consider  the mollifier $F_{n,\theta,\Delta} \cd$ given
by \bea \ad F_{n, \theta ,\Delta} (\phi; u(i \e), i< n, w(p \theta),
p \theta \le n \e) \\
\ad  = N(\Delta) \int \cdots \int F_{n,\theta} (\phi; u(i
\e), i<n,
w( p \theta) + z_p, p \theta \le n \e) \times \prod_p \exp (- | z_p|^2
/(2\Delta)) d
z_p,\eea where $N(\Delta)$ is a normalizing constant so the integral
of the mollifier is unity. Note that $F_{n, \theta ,\Delta}$ are
nonnegative, and they are continuous in the $w$-variable. As $
\Delta \to 0$, $F_{n,\theta,\Delta}$ converges to $F_{n, \theta}$ with
probability one.
Let $ u^{\varsigma, \theta, \Delta}\cd$ be the piecewise constant admissible
control that is determined by the conditional probability
distribution $F_{n,\theta ,\Delta}\cd$. There is a probability space on
which we can define $w^{\varsigma, \theta ,\Delta}\cd$ and the control
law $u^{\varsigma, \theta, \Delta}\cd$ by the conditional probability \bea
\ad P( u^{\varsigma,\theta,\Delta} (n \e)=\phi | u^{\varsigma ,\theta,\Delta}(i
\e), i< n, w^{\varsigma ,\theta ,\Delta}(\tau), \tau \le n \e)\\
\aad \ =F_{n,\theta ,\Delta} (\phi; u^{\varsigma \theta, \Delta} (i \e), i< n,
w^{\varsigma, \theta ,\Delta}( p \theta), p \theta \le n\e).\eea Then
$(x^{\varsigma,\theta,\Delta}\cd, \alpha^{\varsigma,\theta,\Delta}\cd,
m^{\varsigma,\theta,\Delta}\cd, w^{\varsigma,\theta,\Delta}\cd)$ converges
weakly to $(x^{\varsigma,\theta}\cd, \alpha^{\varsigma,\theta},
m^{\varsigma,\theta}\cd,
w^{\varsigma,\theta}\cd)$  as $\Delta\to 0$. This yields that
$| J^{m^{\varsigma, \theta,\Delta}}_{x,\iota}\cd -
J^{m^{\varsigma ,\theta}}_{x,\iota}\cd| \le g_2 (\Delta)$,
where
$g_2(\Delta)\to 0$ as $\Delta\to 0$.

Finally, choose $\wdt \varsigma$ sufficiently small. Then for each
$\varsigma>0$, there are $\e>0$, $\theta>0$, $w^\varsigma\cd$, and an
admissible control that is piecewise constant on $[n \e,n\e+\e)$
taking values in a finite set $U^\varsigma\subset U$ determined by the
conditional probability law \bea \ad P( u^\varsigma ( n\e)=\phi| u^\varsigma(
i\e), i<n, w^\varsigma(\tau), \alpha^\varsigma(\tau), \tau \le n \e)\\ \aad \ \   =
F_n (\phi; u^\varsigma(i\e), i< n, w^\varsigma(p\theta),
\alpha^\varsigma(p\theta),
p \theta\le n \e),\eea where $F_n\cd$ are continuous w.p.1 in the
$w$-variables for each of other variables. Moreover,
(\ref{diff-www}) holds. \qed

\end{document}